\documentclass[12pt,letterpaper]{article}
\usepackage{jheppub}

\usepackage{graphicx}
\usepackage{bbm}
\usepackage{amsmath}
\usepackage{amssymb}
\usepackage{mathtools}
\usepackage{amsthm}

\usepackage[labelformat=simple]{subcaption}

\newcommand{\cM}{\mathcal{M}}

\newcommand{\unit}[1]{\hat{\mathrm{\bold{#1}}}}
\def\be{\begin{equation}}
\def\ee{\end{equation}}

\newtheorem{Definition}{Definition}

\newtheorem{NewConjecture}{New Conjecture}

\newtheorem{ToyConjecture}{Toy Conjecture}

\newtheoremstyle{named}{0.75\baselineskip}{0.75\baselineskip}{\itshape}{}{\bfseries}{.}{.5em}{#3}
\theoremstyle{named}
\newtheorem*{namedconjecture}{Conjecture}
\newtheorem*{prominentconjecture}{Conjecture}

\newcommand{\Mpld}{M_{{\rm Pl;}d}}

\begin{document}

\begin{titlepage}

{\flushright ACFI-T23-04 \\ }

\setcounter{page}{1} \baselineskip=15.5pt \thispagestyle{empty}

\bigskip\

\vspace{2cm}
\begin{center}
{\LARGE \bfseries Dense Geodesics, Tower Alignment, and the Sharpened Distance Conjecture}

 \end{center}
\vspace{0.5cm}

\begin{center}
{\fontsize{14}{30}\selectfont Muldrow Etheredge}
\end{center}

\begin{center}
\vspace{0.25 cm}
\textsl{Department of Physics, University of Massachusetts, Amherst, MA 01003 USA}\\

\vspace{0.25cm}

\end{center}
\vspace{1cm}
\noindent

The Sharpened Distance Conjecture and Tower Scalar Weak Gravity Conjecture are closely related but distinct conjectures, neither one implying the other. Motivated by examples, I propose that both are consequences of two new conjectures: 1. The infinite distance geodesics passing through an arbitrary point $\phi$ in the moduli space populate a dense set of directions in the tangent space at $\phi$. 2. Along any infinite distance geodesic, there exists a tower of particles whose scalar-charge-to-mass ratio ($-\nabla \log m$) projection everywhere along the geodesic is greater than or equal to $1/\sqrt{d-2}$. I perform several nontrivial tests of these new conjectures in maximal and half-maximal supergravity examples. I also use the Tower Scalar Weak Gravity Conjecture to conjecture a sharp bound on exponentially heavy towers that accompany infinite distance limits.

\vspace{.9cm}

\bigskip
\noindent\today

\end{titlepage}

\setcounter{tocdepth}{2}

\hrule
\tableofcontents

\bigskip\medskip
\hrule
\bigskip\bigskip

\section{Introduction\label{s.introduction}}

In string theory, the parameters are given by vacuum expectation values of massless scalar fields called moduli, and these vevs parametrize manifolds called moduli spaces of string vacua. Charting and studying the properties of these moduli spaces is of great importance in string phenomenology and the Swampland Program \cite{Vafa:2005ui,vanBeest:2021lhn,Agmon:2022thq,Palti:2019pca}.

These moduli spaces have natural metrics. To see this, suppose that the coordinates of a moduli space are given by the vacuum expectation values of some set of moduli $\phi^i$, indexed by $i$. Then\footnote{In this paper, I will consider only the cases where the moduli have no potentials.} the metric on this moduli space is defined by the kinetic matrix $G_{ij}(\phi)$ appearing in the low-energy effective action,
\begin{align}\label{e.gravity_with_massless_scalars}
S=\frac 1{2\kappa_d^2}\int  d ^dx \sqrt{-g}\left(\frac{R}2-\frac12 G_{ij}(\phi) \partial_\mu\phi^i\partial^\mu\phi^j+\dots\right).
\end{align}

With such metrics, significant and well studied conjectures have been made about moduli spaces and the associated spectra. In particular, the seminal paper \cite{Ooguri:2006in} by Ooguri and Vafa proposed several influential conjectures. For instance, given any point $\phi_1$ in a moduli space and any positive real number $T$, Ooguri and Vafa conjectured that one can find a point $\phi_2$ such that the distance between $\phi_1$ and $\phi_2$ is greater than $T$. Building on this, these authors then proposed the following conjecture constraining the spectra of asymptotically far out regions of moduli spaces: 
\begin{prominentconjecture}[The Distance Conjecture]
Let $\cM$ be the moduli space of a quantum gravity theory in $d \geq 4$ dimensions, parametrized by vacuum expectation values of massless scalar fields. Fixing a point $\phi_1 \in \mathcal{M}$, the theory at a point $\phi_2 \in \mathcal{M}$ sufficiently far away in the moduli space has an infinite tower of light particles, with characteristic mass in Planck units $(\kappa_d^2 = \Mpld^{2-d} = 1)$ scaling as 
\be
m(\phi_2) \sim \exp( -\alpha \,||\phi_1,\phi_2|| )\quad \text{as}\quad ||\phi_1, \phi_2||\rightarrow \infty ,
\label{DCdef}
\ee 
where $||\phi_1, \phi_2||$ is the length of the shortest geodesic in $\mathcal{M}$ between $\phi_1$ and $\phi_2$, and $\alpha>0$ is some order-one number.
\end{prominentconjecture}
\noindent

A primary aim of this current paper is to connect the Distance Conjecture with local, rather than asymptotic, statements about moduli spaces. As will be shown below, this can be achieved by introducing two new conjectures about geodesics and towers of particles that together imply a sharpened version of the Distance Conjecture as well as the Tower Scalar Weak Gravity Conjecture (defined later in the introduction), which is a local statement about tangent spaces of moduli space.

The Distance Conjecture has been tested, explored, and refined in a variety of contexts and ways (see, e.g., \cite{Klaewer:2016kiy, Baume:2016psm, Blumenhagen:2017cxt, Grimm:2018ohb, Blumenhagen:2018nts, Grimm:2018cpv, Corvilain:2018lgw, Joshi:2019nzi, Marchesano:2019ifh, Font:2019cxq, Erkinger:2019umg, Buratti:2018xjt, Heidenreich:2018kpg, Gendler:2020dfp, Lanza:2020qmt, Klaewer:2020lfg, Rudelius:2023mjy}). In \cite{Etheredge:2022opl}, a sharp lower bound on the exponential rate for the lightest tower in the Distance Conjecture was first proposed, resulting in the Sharpened Distance Conjecture:
\begin{namedconjecture}[The Sharpened Distance Conjecture]
The Distance Conjecture remains true with the added requirement that
\begin{align}
\alpha \geq \frac{1}{\sqrt{d-2}},
\end{align}
where $d$ is the spacetime dimension.
\end{namedconjecture}
\noindent The Sharpened Distance Conjecture has been tested in various contexts \cite{Etheredge:2022opl, Etheredge:2023odp} and will be the primary focus of this paper.

The Sharpened Distance Conjecture is closely connected to, and perhaps implied by, the Emergent String Conjecture proposed in \cite{Lee:2019wij,Lee:2019xtm}. The Emergent String Conjecture states that infinite distance limits are either decompactification limits or limits in which fundamental strings become tensionless, and this has been tested in a variety of contexts \cite{Lee:2018urn,Lee:2019xtm, Lanza:2021udy, Baume:2019sry, Xu:2020nlh,Baume:2020dqd,Perlmutter:2020buo,Baume:2023msm}. The Sharpened Distance Conjecture seems related to the Emergent String Conjecture because, in tensionless string limits, the string oscillators of perturbative strings have $\alpha$'s satisfying $\alpha^\text{osc} =1/\sqrt{d-2}$, whereas in all studied cases thus far decompactification limits have KK-modes with coefficients $\alpha^\text{KK}\geq  1/\sqrt{d-2}$. However, for KK modes, it is not yet clear whether $\alpha^\text{KK}\geq1/\sqrt{d-2}$ is always true. For typical toroidal decompactifications from $d$-dimensions to $D=d+n$ dimensions, the formula for the KK-modes is
\begin{align}
	\alpha^\text{KK} =\sqrt{\frac{D-2}{n(d-2)}}\geq \frac 1{\sqrt{d-2}}.\qquad\text{(often)}\label{e.alphanaive}
\end{align}
A general formula for KK modes when running decompactification occurs is not currently known, and there are examples of decompactification \cite{Etheredge:2023odp} where the formula \eqref{e.alphanaive} does not apply. So, it is not clear whether the Sharpened Distance Conjecture follows from the Emergent String Conjecture, unless one can be sure that $\alpha^\text{KK}\geq 1/\sqrt{d-2}$. It is also not fully clear what happens in emergent string limits that are not strictly perturbative.

Let us turn our attention to scalar charges, which are sections of the tangent bundles of moduli spaces. For two particles of mass $m_1$ and $m_2$ separated by a long distance, the $1/r^{d-2}$ component of the long-range force between them is proportional to \cite{Heidenreich:2019zkl,Heidenreich:2020upe}
\begin{align}
	\mathcal F_{12}=f^{AB}Q_{1A}Q_{2B}- G^{ab}\mu_{1a}\mu_{2b}-\frac{d-3}{d-2}\kappa_d^2 m_1 m_2, \label{e.lrf}
\end{align}
where $Q_{i,A}$ is the charge of the $i$th particle under the $A$th U$(1)$ gauge field, $f^{AB}$ is the inverse gauge kinetic matrix, $G^{ab}$ is the inverse scalar kinetic matrix, and $\mu_{ia}$ is $a$-th component of the scalar charge of the $i$th particle. The scalar charge for a particle of moduli-dependent mass $m(\phi)$ is defined as the negative gradient (with respect to the moduli) of the mass of that particle,
\begin{align}
	\vec\mu\equiv-\nabla m.
\end{align}
The reason that $\vec \mu$ is called scalar charge is because it appears in the long range force formula \eqref{e.lrf} in a way that is analogous to the appearance of the electric charge. However, the word ``charge" should not be taken to imply conserved, because scalar charges are not conserved, unlike electric and magnetic charges.

It is convenient to rescale scalar charges and introduce ``$\alpha$-vectors":
\begin{Definition}[Scalar-charge-to-mass ratio ($\alpha$-vector)]
	Consider a particle with moduli-dependent mass $m(\phi)$. Then the \textbf{scalar-charge-to-mass ratio}, or ``$\alpha$-vector", for this particle is defined as
	\begin{align}
		\vec{\alpha}=-\nabla \log m,
	\end{align}
	where the gradient is with respect to the moduli and the Planck mass is set to one.
\end{Definition}

These $\alpha$-vectors allow for a local conjecture about the tangent bundle of moduli space called the Tower Scalar Weak Gravity Conjecture (Tower SWGC) \cite{ Calderon-Infante:2020dhm, Etheredge:2022opl, Etheredge:2023odp}:
\begin{namedconjecture}[Tower Scalar Weak Gravity Conjecture (Tower SWGC)]
	Consider an arbitrary point $\phi$ in the moduli space $\mathcal M$. Suppose that there exists a set of towers of particles at that point. The closure of the convex hull generated by the set of $\alpha$-vectors for these towers contains the ball of radius $1/\sqrt{d-2}$, where $d$ is the number of spacetime dimensions.
\end{namedconjecture}
\noindent The Scalar Weak Gravity Conjecture (SWGC) has appeared in a variety of contexts and has meant different things in different contexts, but the above version is the version I will use in this paper.\footnote{Originally, several papers investigated inequalities involving scalar forces (see, e.g.,  \cite{Palti:2017elp, Lee:2018spm, Andriot:2020lea, Gonzalo:2019gjp, Freivogel:2019mtr, DallAgata:2020ino, Benakli:2020pkm, Gonzalo:2020kke}). In \cite{Calderon-Infante:2020dhm}, the authors introduced a convex hull version of the SWGC, but without fully specifying the region enclosed in the convex hulls. Eventually in \cite{Etheredge:2022opl}, it was argued that the convex hull of $\alpha$-vectors contains a ball centered at the origin with radius at least $1/\sqrt{d-2}$. }

The version of the Tower SWGC used in this paper is subtly but importantly different from the version stated in \cite{Etheredge:2022opl,Etheredge:2023odp}. The version in \cite{Etheredge:2022opl,Etheredge:2023odp} states that the convex hull of $\alpha$-vectors contains the ball of radius $1/\sqrt{d-2}$, but the correct version of the Tower SWGC holds that the convex hull of the \emph{closure} of the set of $\alpha$-vectors that contains the ball of radius $1/\sqrt{d-2}$. Sometimes there is an uncountably infinite set of points on the boundary of the ball of radius $1/\sqrt{d-2}$ that \emph{are not} in the convex hull of $\alpha$-vectors. This phenomenon occurs in, for instance, IIB string theory in 10d and in 9d, and this will play an important role in sharpening the connections between the Tower SWGC, geodesics, and the Distance Conjecture.

The Tower SWGC has been connected\footnote{In Section \ref{s.conjectures} of this paper, I propose several refinements of the Convex Hull Distance Conjecture of \cite{Calderon-Infante:2020dhm}.} with the Distance Conjecture \cite{Palti:2017elp, Gendler:2020dfp, Calderon-Infante:2020dhm, Grimm:2018ohb, Grimm:2022sbl, Etheredge:2022opl, Etheredge:2023odp}. But, as will be demonstrated in this paper, such a relationship is complicated. This is because the Distance Conjecture requires that $\alpha$-vectors align with infinite distance limits, but $\alpha$-vectors can behave in complicated ways. As demonstrated in \cite{Etheredge:2022opl,Etheredge:2023odp}, $\alpha$-vectors often ``slide" (i.e., have moduli-dependent lengths and directions). Even worse, the existence of certain towers can be moduli-dependent.

The main focus of this paper is to illustrate how, in maximal supergravity, the $\alpha$-vectors conspire with geodesics in ways that imply both the Tower SWGC and the Sharpened Distance Conjecture. I find in examples that the following two new conjectures hold and, when combined, imply both the Sharpened Distance Conjecture and the Tower SWGC. I propose the following two new conjectures for maximal supergravity theories (and possibly more theories).

\begin{NewConjecture}[Dense Direction Conjecture]
	Consider an arbitrary point $\phi$ in the moduli space and all geodesics passing through $\phi$ that go to infinite distance limits. The set of directions of these geodesics at $\phi$ is dense in the set of all directions in the tangent space at $\phi $.
\end{NewConjecture}

\begin{NewConjecture}[Tower Alignment Conjecture]
	Consider an arbitrary infinite distance geodesic $\gamma$. Then there exists a tower with an $\alpha$-vector such that its projection along the geodesic satisfies, at all points along the geodesic,
	\begin{align}
	\alpha_\parallel\geq 1/\sqrt{d-2}.
	\end{align}
\end{NewConjecture}
\noindent I also discuss a possible weakening of this conjecture where the towers only are required to exist and align when their masses are light enough, perhaps being below the species scale.

These two new conjectures combine to imply the Tower SWGC and the Sharpened Distance Conjecture. In this paper, I provide tests of these conjectures where these two new conjectures just barely manage to pass. In particular, I highlight examples where $\alpha$-vectors of length $1/\sqrt{d-2}$ just barely manage to align everywhere with non-straight geodesics in curved spaces. Such examples provide nontrivial tests of these conjectures.

In addition to the above two conjectures, I also discuss how the Tower SWGC motivates the following new conjecture about heavy towers appearing in infinite distance limits:
\begin{NewConjecture}[Heavy Towers Conjecture]
	Fixing a point $\phi_1$ in the moduli space, the theory at a sufficiently far away point $\phi_2$ in the moduli space has an infinite tower of heavy particles with characteristic mass scaling as
	\begin{align}
		m\sim e^{|\alpha|||\phi_1,\phi_2||},\qquad \text{as}\qquad ||\phi_1,\phi_2||\rightarrow\infty,
	\end{align}
	where $||\phi_1,\phi_2||$ is the length of the shortest geodesic in $\mathcal M$ between $\phi_1$ and $\phi_2$, and $|\alpha|\geq 1/\sqrt{d-2}$.
\end{NewConjecture}
This proposal implies that under any infinite-distance limit, there exists a tower that becomes exponentially heavy. While such towers in some contexts may be anticipated by duality, this proposal places a precise lower bound on the exponential rate of the heavy towers. Heavy towers have also been observed in \cite{Ahmed:2023cnw}.

In this paper, I perform explicit tests of these conjectures in 10d and 9d maximal supergravity cases. For 16 supercharge cases in 9d, I perform limited, but nontrivial, tests of these conjectures in asymptotic regions of the moduli space.

One might worry about the Tower SWGC when the necessary towers are not BPS, because these non-BPS towers may be unstable when heavy, or ill-defined in non-perturbative regimes. This is not a problem when considering M-theory reduced on $T^2$ through $T^7$ (and possibly also $T^8$) \cite{Etheredge:2023odp} since in these cases the conjecture is satisfied by considering only 1/2 BPS particles. But, the Tower SWGC is confronted with this issue in 10d IIA and IIB string theory, since the oscillator modes of the strings are necessary for the Tower SWGC to hold in strong-coupling regimes.

To deal with the non-BPS states in 10d maximal supergravity cases, it is interesting to consider a closely-related stringy alternative to the Tower SWGC that does not need to involve heavy non-BPS towers in strongly coupled regimes. As will be discussed below, I will use this alternative as a proxy for testing the regular Tower SWGC in 10d maximal supergravity examples. Before I define this Stringy SWGC, I first generalize the notion of scalar-charge-to-mass ratios, or $\alpha$-vectors, from particles to branes.
\begin{Definition}[Scalar-charge-to-tension ratio for branes ($\alpha$-vector)]
	Consider a $p$-brane with moduli-dependent tension $T(\phi)$. Then the \textbf{scalar-charge-to-tension ratio}, or ``$\alpha$-vector", for this brane is defined as
	\begin{align}
		\vec{\alpha}=-\frac 1{p+1}\nabla \log T,
	\end{align}
	where the gradient is with respect to the moduli and the Planck mass is set to one.
\end{Definition}
Importantly, this definition does not require weak coupling, and so it makes sense in strong coupling regimes, where BPS branes may be present but oscillator modes might be difficult to study. When $p=0$, this definition describes an $\alpha$-vector for a particle. When $p=1$, this definition produces the $\alpha$-vector for a string oscillator tower if the masses scale with the square-root of the string tension, as is the case for fundamental strings in low-tension limits. But, such oscillators are not required for this generalized definition of $\alpha$-vectors. The power of this definition is that it allows us to work with $\alpha$-vectors for strings (which are often well-defined and BPS objects), avoids speculation about nonperturbative properties of string oscillators, and still constrains the $\alpha$-vectors for these oscillators in tensionless limits.\footnote{In this paper, I consider only ``fundamental strings" that oscillate with towers in low tension limits, and I do not consider cases where low-tension strings do not oscillate with towers. For more on this distinction, see \cite{Reece:2018zvv}.}

In this paper, I show that the following stringy version of the SWGC holds in the maximal supergravity examples examined in this paper.
\begin{NewConjecture}[Stringy Scalar Weak Gravity Conjecture (Stringy SWGC)]
	Consider an arbitrary point $\phi$ in the moduli space $\mathcal M$.  \textbf{Stringy Scalar Weak Gravity Conjecture} requires that the following two conditions are met at $\phi$.
	\begin{enumerate}
	\item There exists a set of towers of particles and/or there also exists some set of fundamental strings.
	\item The completion of the convex hull generated by the set of $\alpha$-vectors for these towers and strings contains the ball of radius $1/\sqrt{d-2}$, where $d$ is the number of spacetime dimensions.
	\end{enumerate}
\end{NewConjecture}
\noindent In this paper, I discuss evidence for this Stringy SWGC in maximal supergravity examples, and I also discuss tests of it in half-maximal supergravity examples.

The Stringy SWGC and the Tower SWGC are closely related, but the Stringy SWGC does not necessarily imply the Tower SWGC. In the 10d maximal supergravity examples in this paper, I will use tests of the Stringy SWGC as substitutes for directly testing the Tower SWGC. This is justified in weak-coupling regimes, where string oscillation tower masses scale with the square roots of the tensions of the strings. In these regimes the Stringy SWGC implies the Tower SWGC. However, this approach is subtle in strong coupling regimes. I leave it to future work to see to what extent the Stringy SWGC can tell us about the Tower SWGC.

We can explore direct connections between the Stringy (and Tower) SWGC and the Distance Conjecture with the following example. Suppose one is traveling an infinite distance limit given by a unit-speed geodesic $\gamma(t)$, and suppose that the $\alpha$-vector for a string or particle tower satisfies everywhere on that geodesic (by the Tower Alignment Conjecture)
\begin{align}
	\vec \alpha \cdot \dot \gamma\geq \frac{1}{\sqrt{d-2}}.\label{e.proj}
\end{align}
Then this string, or tower, gets a tension or mass that scales asymptotically with the distance
\begin{align}
	T_p\lesssim e^{-\frac {p+1}{\sqrt{d-2}}t}, \label{e.tensionless}
\end{align}
with $p=0$ for the particle case and $p=1$ for the string case. For the case where \eqref{e.proj} is satisfied by a particle tower, the Sharpened Distance Conjecture is satisfied in this limit since the equation \eqref{e.tensionless} is precisely the exponential behavior required by the Sharpened Distance Conjecture.\footnote{Note that, if we further require in the Stringy SWGC that these towers have KK-like tower spacing $m_n\sim n$, then these towers resemble KK towers of decompactification limits, thus resembling the predictions of the Emergent String Conjecture.} Meanwhile, for the case where \eqref{e.proj} is satisfied by a fundamental string, we obtain a weak-coupling limit of the string as the string becomes tensionless. In this low-tension regime, there is a tower of string oscillators that scale with the square root of the tension. So, in this regime, equation \eqref{e.tensionless} tells us that the tower of string oscillators gives us an emergent string limit. Thus, the Stringy SWGC can imply the Sharpened Distance Conjecture, when accompanied by certain additional assumptions that this paper makes precise. This example demonstrates the essence of what happens in the examples considered in this paper. For non-flat moduli spaces, this phenomenon requires highly-nontrivial alignment properties between $\alpha$-vectors and infinite-distance geodesics, as formalized by the Tower Alignment Conjecture. 

It should be emphasized that neither the Stringy SWGC nor the Tower SWGC implies the Distance Conjecture without additional assumptions. The perspective proposed by this paper is that both the Tower SWGC and Sharpened Distance Conjecture can be viewed as consequences of the Tower Alignment and Dense Direction Conjectures being simultaneously true. It is these two new conjectures that precisely connect the Sharpened Distance Conjecture and Tower SWGC.

An outline of this paper is as follows. In Section \ref{s.sometimes}, I examine how in special examples the Tower SWGC implies the Sharpened Distance Conjecture and the Heavy Towers Conjecture. This involves examining a toy example of the Tower SWGC where the $\alpha$-vectors are moduli independent. In Section \ref{s.failure}, I discuss how in general the Tower SWGC does not imply the Sharpened Distance Conjecture, as can when $\alpha$-vectors are moduli dependent and slide. In Section \ref{s.conjectures} I propose new conjectures. In particular, when the Tower Alignment and Dense Direction Conjectures are both true, the Sharpened Distance Conjecture and Tower SWGC both follow as consequences. In Sections \ref{s.IIB10d} and \ref{s.329d}, I show\footnote{For the 10d case, I test the Stringy SWGC as a proxy for the Tower SWGC.} how these conjectures are satisfied in 10d and 9d theories with maximal supergravity, and in Section \ref{s.169d} I examine these conjectures in 9d theories with half-maximal supergravity. I conclude in Section \ref{s.summary} with a summary and list of future directions.

\section{How the Tower SWGC sometimes implies the Distance Conjecture \label{s.sometimes} }

In this section, I discuss toy examples where the Stringy SWGC, and Tower SWGC, imply the Sharpened Distance Conjecture. However, in Section \ref{s.failure}, I show that the Tower SWGC does not imply the Sharpened Distance Conjecture in general.

\begin{figure}
\begin{center}
\includegraphics[width = 80mm]{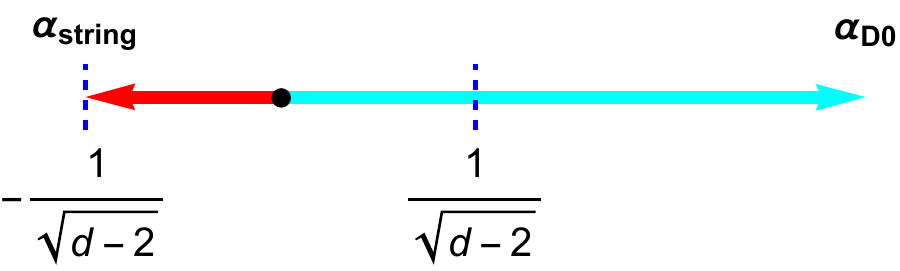}
\end{center}
\caption{$\alpha$-vectors of fundamental strings and D0-branes in 10d IIA string theory. The blue dashes indicate the ball of radius $1/\sqrt{d-2}$ centered at the origin.}
\label{f.IIA10dalphas}
\end{figure}

A simple first example to consider is IIA string theory in 10d. Here, the moduli space is one dimensional, and the Stringy SWGC is satisfied by D0-branes and fundamental strings. These states have $\alpha$-vectors satisfying
\begin{align}
	\alpha_\text{D0}=\sqrt{\frac{9}{8}},\qquad\alpha_\text{string}=-\frac{1}{\sqrt 8}.
\end{align}
These are depicted in Figure \ref{f.IIA10dalphas}.

We can use the Stringy SWGC to imply the Sharpened Distance Conjecture. Consider a point $\phi_1$ in the moduli space. At $\phi_1$, we have that the D0-branes have a tower with characteristic mass $m_\text{D0}(\phi_1)$ and we have that the fundamental strings have some tension $T_\text{str}(\phi_1)$. Consider a different point $\phi_2$. We can use the $\alpha$-vectors of the D0-branes and strings to relate the masses and tensions at $\phi_2$ with the masses and tensions at $\phi_1$. For the D0-branes and strings, these relations are given by
\begin{align}
	\log \frac{m_\text{D0}(\phi_2)}{m_\text{D0}(\phi_1)}&=\int_{\phi_1}^{\phi_2}\frac{d \log m_\text{D0}}{d\phi}d\phi =-\alpha_\text{D0}\times (\phi_2-\phi_1),\\
	\log \frac{T_\text{string}(\phi_2)}{T_\text{string}(\phi_1)}&=\int_{\phi_1}^{\phi_2}\frac{d \log T_\text{string}}{d\phi}d\phi =-2\alpha_\text{string}\times (\phi_2-\phi_1).
\end{align}
So,
\begin{align}
	m_\text{D0}(\phi_2)=m_\text{D0}(\phi_1)e^{-\sqrt{\frac{9}{8}}(\phi_2-\phi_1)},\qquad T_\text{string}(\phi_2)=T_\text{string}(\phi_1)e^{\frac{2}{\sqrt{8}}(\phi_2-\phi_1)}. 
\end{align}
In this case, the Sharpened Distance Conjecture follows from the Stringy SWGC. To see this, suppose we take the $\phi_2\gg\phi_1$ limit. Then the D0-brane masses become exponentially light, providing a decompactification KK tower. If we instead take the $\phi_2\ll \phi_1$ limit, the string becomes exponentially weakly coupled, and we get an exponentially light tower of string oscillation modes satisfying the Sharpened Distance Conjecture bound.

One might hope that the Tower SWGC, and not just the Stringy SWGC, applies here and that it also implies the Sharpened Distance Conjecture. However, the Tower SWGC holding everywhere in this example requires string oscillators in strongly-coupled regimes. This is subtle since such oscillators are non-BPS. Nevertheless, let us momentarily suppose such towers exist and that their dependence on the moduli is given by
\begin{align}
	m_\text{osc}\sim \sqrt{T_\text{string}}.\label{e.squareroottension}
\end{align}
With these assumptions, the Tower SWGC is satisfied everywhere in moduli space, since the $\alpha$-vectors of these oscillators and the D0 branes generate the requisite convex hull. Also, with these assumptions, the Tower SWGC implies the Sharpened Distance Conjecture, due to an argument similar to the argument above where the Stringy SWGC was used to imply the Sharpened Distance Conjecture.

There is another interesting consequence from assuming \eqref{e.squareroottension}, and hence the Tower SWGC, in this example. If we go in strong coupling regimes, the oscillators become heavy. If we go in weak coupling regimes, the D0 branes become heavy. More precisely, the Tower SWGC suggests that every infinite distance limit is not only accompanied with a tower of light particles, but also with a tower of heavy particles that become exponentially heavy, with masses scaling with
\begin{align}
	m(\phi)\sim e^{\alpha||\phi_1,\phi_2||},
\end{align}
where $\alpha\geq 1/\sqrt{d-2}$.

Before moving on to higher dimensional moduli spaces, I re-emphasize that in strong-coupling regimes, it is not clear whether \eqref{e.squareroottension} holds, and thus whether the Tower SWGC follows from the Stringy SWGC. This will be the subject of future work. In this paper, we will content ourselves with sometimes using only the Stringy SWGC and not the Tower SWGC, as we have done in this 10d IIA string theory example. This approach will again have to be taken in the 10d IIB string theory example. Fortunately, in the 9d, 8d, 7d, 6d, 5d, 4d (and possibly 3d) maximal supergravity cases, the Tower SWGC is actually satisfied by BPS particles \cite{Etheredge:2022opl}. My 9d maximal supergravity analysis in Section \ref{s.329d}, as well as my analysis in 9d half-maximal supergravity in Section \ref{s.169d}, will not rely on the Stringy SWGC.

Let us next consider higher-dimensional moduli spaces. It is tempting to extrapolate the above reasoning and claim that the Tower SWGC implies the Sharpened Distance Conjecture for such moduli spaces, but this is not true without further assumptions. For instance, the moduli space of M-theory on $T^2$ is three-dimensional and the $\alpha$-vectors depend on the moduli and slide, invalidating the above approach taken for the 10d IIA case.

\begin{figure}
\begin{center}
\includegraphics[width = 80mm]{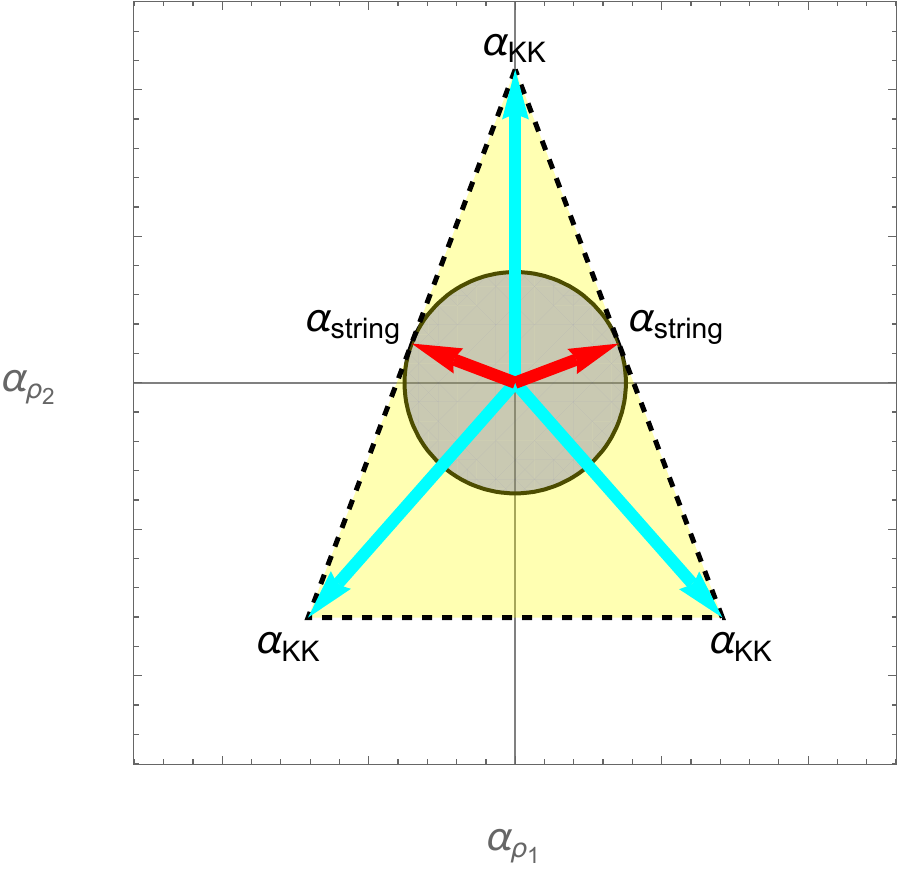}
\end{center}
\caption{Radion-radion components of M-theory on $T^2$.}
\label{f.IIA9dslice}
\end{figure}

Nevertheless, the situation simplifies considerably if we consider only the radion-radion slice of the moduli space of M-theory on a two-torus, as depicted in Figure \ref{f.IIA9dslice}.
In this example, the $\alpha$-vectors on the vertices of this triangle are from 1/2 BPS particles and do not change as we move around the radion-radion slice of moduli space. If one travels in any infinite distance limit in radion-radion moduli space, the Tower SWGC implies the Sharpened Distance Conjecture. In this example, one must be careful to only move in the radion-radion plane and not vary the axion---that is, one must be careful to change the radii of the cycles of the torus without changing the shape of the torus. As we will see later, the 1/2 BPS states in this example can have moduli-dependent $\alpha$-vectors if we change the angle between the two torus cycles.

The 10d IIA example and the radion-radion slice of M-theory on $T^2$ inspire a toy version of the Tower SWGC that implies the Sharpened Distance Conjecture. Consider the following toy version of the Tower SWGC where the $\alpha$-vectors are held constant and do not slide as one moves around the moduli space:
\begin{ToyConjecture}[Non-Sliding Tower SWGC]
 The Tower SWGC is satisfied for a moduli space with globally flat coordinates and the $\alpha$-vectors that generate the convex hull are moduli-independent.
\end{ToyConjecture}
This toy example is interesting because it applies to some examples, implies the Sharpened Distance Conjecture, and also implies the existence of heavy towers in asymptotic limits in the moduli space.

Let us see explicitly how this toy conjecture implies the Sharpened Distance Conjecture. Suppose we have two points $\phi_1$ and $\phi_2$, and suppose that $\gamma$ is a geodesic connecting these points. Then the non-sliding Tower SWGC implies that there is a particle tower with an $\alpha$-vector such that its projection along the geodesic $\gamma$ satisfies $\alpha_\parallel\geq 1/\sqrt{d-2}$. If this tower has a characteristic mass $m_1$ at the point $\phi_1$, then the characteristic mass at the point $\phi_2$ is bounded by the characteristic mass at $\phi_1$ by
\begin{align}
\begin{aligned}
	\log \frac{m_2 }{m_1}&=\int_{\phi_1}^{\phi_2}(\nabla \log m)\cdot d\textbf{l}=-\int_{\phi_1}^{\phi_2}\alpha_\parallel dl\leq -\frac 1{\sqrt{d-2}}||\phi_2,\phi_1||,\\
	\Rightarrow \qquad m_2&\leq m_1 e^{-\frac 1{\sqrt{d-2}}||\phi_2,\phi_1||},\label{e.naiveparticletoy}
\end{aligned}
\end{align}
thus implying the Sharpened Distance Conjecture in this geodesic. If this tower has a tower spacing of a KK tower, i.e. $m_n\sim n$, then this tower behaves like a KK tower in a decompactification limit.

One can also consider a similar toy version of the Stringy SWGC:
\begin{ToyConjecture}[Non-Sliding Stringy SWGC]
 The Stringy SWGC is satisfied for a moduli space with globally flat coordinates and the $\alpha$-vectors that generate the convex hull are moduli-independent.
\end{ToyConjecture}
The Sharpened Distance Conjecture is also a consequence of the Non-Sliding Stringy SWGC. For the case where there is a particle tower along the geodesic, the argument above from the Non-Sliding Tower SWGC implies the Sharpened Distance Conjecture. For the case where instead there is just a string with an $\alpha$-vector aligned with the geodesic, then an analogous argument implies that the tension at $\phi_2$ is bounded by the tension at $\phi_1$ by
\begin{align}
	T_2\leq T_1 e^{-\frac{2}{\sqrt{d-2}}||\phi_2,\phi_1||}.
\end{align} 
If $\phi_2$ and $\phi_1$ are sufficiently separated, then $T_2$ is low-tension, and there is a light tower of oscillators with characteristic mass equal to the square root of the tension, and this results in the characteristic mass of this tower having an exponential scaling that satisfies the Sharpened Distance Conjecture.

The Non-Sliding Tower SWGC inspires conjectures about heavy towers. If we have the Non-Sliding Tower SWGC, then there is a formula analogous  to \eqref{e.naiveparticletoy} but with a different sign in the exponent. This is because along geodesics the convex hull condition implies that there are towers with $\alpha$-vectors that are anti-aligned with the geodesic, as opposed to the light towers that have aligned $\alpha$-vectors. That is, the convex hull condition of the Non-Sliding Tower SWGC implies that there exist towers with $\alpha_\parallel\leq -1/\sqrt{d-2}$, thus implying that there exist towers also with relations
\begin{align}
	\log \frac{m_b }{m_a}&=\int_a^b(\nabla \log m)\cdot d\textbf{l}=-\int_a^b\alpha_\parallel dl\geq \frac 1{\sqrt{d-2}}||b,a||,\\
	\Rightarrow \qquad m_b&\geq m_a e^{\frac 1{\sqrt{d-2}}||b,a||}.
\end{align}
Such heavy towers also potentially follow from the Non-Sliding Stringy SWGC, where the towers are related to oscillators of the high-tension strings, but these towers will be investigated more thoroughly in future work.

These toy conjectures are not true in general. Almost all moduli spaces are curved\footnote{The only flat moduli spaces that I know of either occur in one-dimensional moduli spaces, such as the moduli space of IIA string theory in 10d, or on carefully obtained submanifolds of moduli space.} and so these spaces do not have globally flat coordinates, and global flatness is required to be able to discuss the convex hull of $\alpha$-vectors globally. Also, in many examples the $\alpha$-vectors are moduli-dependent and slide. However, sometimes carefully obtained submanifolds of moduli space and slices of convex hulls of $\alpha$-vectors satisfy this toy conjecture. A partial classification of such non-sliding examples on flat moduli spaces/slices will appear in \cite{Etheredge:Taxonomy}.

\section{How the Tower SWGC can fail to imply the Distance Conjecture \label{s.failure}}

If the $\alpha$-vectors in the Tower SWGC are moduli-dependent and slide, the Distance Conjecture is not implied by the Tower SWGC. To see this, suppose that the Tower SWGC is satisfied by some set of $\alpha$-vectors at each point in moduli space. As we move along a geodesic to an infinite distance limit, it may be the case that the $\alpha$-vectors are sometimes aligned with this infinite distance limit, but at other times not aligned. At each point in moduli space, the Tower SWGC can be satisfied, but the Distance Conjecture is not necessarily implied.

Sliding of $\alpha$-vectors is very common. It has been observed both in maximal supergravity \cite{Etheredge:2022opl} and also in 9d theories with 16 supercharges \cite{Etheredge:2023odp}. This sliding even happens with 1/2 BPS states \cite{Etheredge:2022opl}. In the papers \cite{Etheredge:2022opl,Etheredge:2023odp}, sliding is observed in two different contexts. In the maximal supergravity context \cite{Etheredge:2022opl}, axions play a heavy role in the sliding. In the 16 supercharge case \cite{Etheredge:2023odp}, the sliding also occurs during a decompactification to a running solution.

One can see the sliding in maximal supergravity most easily by studying 10d IIB string theory. In 10d IIB string theory, the $\alpha$-vectors for $(p,q)$-strings slide around depending on the location in the moduli space. Additionally, in the M-theory on $T^2$ case, the $\alpha$-vectors for BPS states densely populate a cone \cite{Etheredge:2022opl}, and all of the states on the boundary of this cone, with the exception of the states at the tip, slide around depending on the location on the moduli space. Thus, even in these maximal supergravity examples, the connection between the Tower SWGC and the Distance Conjecture is nontrivial.

In 9d theories with 16 supercharges, there can also be sliding from a running decompactification effect studied in depth in \cite{Etheredge:2023odp}. In this case, there is sliding in this limit in precisely a way so that the Emergent String Conjecture is satisfied.

There is another possibility that does not involve sliding where the Tower SWGC can fail to imply the Distance Conjecture. The Tower SWGC does not imply the Distance Conjecture if the towers whose $\alpha$-vectors allow for the Tower SWGC do not exist everywhere in the moduli space. In the 9d maximal supergravity example studied in this paper, the towers and strings required are BPS states and thus do exist everywhere in moduli space, so this is not an issue for those examples. But this is a possible issue for the 10d maximal supergravity and 9d half-maximal supergravity examples of this paper. In this paper, I focus primarily on either BPS states or non-BPS states in asymptotic limits of moduli space where the states can be trusted to exist and be stable.

\section{New conjectures\label{s.conjectures}}

In this section I discuss several conjectures that simultaneously imply the Tower SWGC and the Sharpened Distance Conjecture. These conjectures are general enough to apply to cases where the moduli spaces are not flat and the $\alpha$-vectors slide.

To make progress, it is useful to precisely define infinite distance geodesics and infinite distance limits.
\begin{Definition}[Infinite Distance Geodesic]
	Let $\phi$ be an arbitrary point in moduli space and let $\gamma$ be a geodesic parameterized by the unit-speed parameter $t$ such that $\gamma(t=0)=\phi$. The geodesic $\gamma$ is an \textbf{infinite distance geodesic} if for any $N>0$ there exists a $T>0$ such that for all $t\geq T$, the length of the shortest path connecting $\phi$ and $\gamma(t)$ is greater than $N$.
\end{Definition}
An infinite distance limit $P_\infty$ can be defined using the following equivalence relation between infinite distance geodesics. 
\begin{Definition}[Infinite Distance Limit\label{d.limit}]
	Two infinite-distance geodesics $\gamma_1$ and $\gamma_2$ are said to give the same infinite distance limit $P_\infty$ if for every $\epsilon>0$ there exists a $T>0$ such that for any $t\geq T$, the distance between $\gamma_1(t)$ and the closest point on $\gamma_2$ is less than $\epsilon$ and the distance between $\gamma_2(t)$ and the closest point on $\gamma_1$ is less than $\epsilon$.
\end{Definition}

Following \cite{Calderon-Infante:2020dhm}, it is useful to rephrase the Sharpened Distance Conjecture in terms of infinite distance limits, geodesics, and $\alpha$-vectors for particle-towers.
\begin{namedconjecture}[Weak Tower Alignment Conjecture]
	Consider an arbitrary point $\phi$ in the moduli space and an arbitrary infinite distance geodesic $\gamma$ passing through $\phi$. Then there exists a tower with an $\alpha$-vector such that its projection along the geodesic satisfies
	\begin{align}
\alpha_\parallel\geq 1/\sqrt{d-2},
	\end{align}
	provided that one is sufficiently far down the geodesic.
\end{namedconjecture}
\noindent This conjecture does not specify how far one has to go down the geodesic to encounter the tower with $\alpha_\parallel \geq 1/\sqrt{d-2}$. The conjecture only states that eventually a tower exists with this property, as this is all that is needed for the Sharpened Distance Conjecture.

Maximal supergravity examples motivate a much stronger version of the Sharpened Distance Conjecture:
\begin{namedconjecture}[Tower Alignment Conjecture]
	Consider an arbitrary infinite distance geodesic $\gamma$. Then there exists a tower with an $\alpha$-vector such that its projection along the geodesic satisfies, at all points along the geodesic,
	\begin{align}
\alpha_\parallel\geq 1/\sqrt{d-2}.
	\end{align}
\end{namedconjecture}
\noindent This differs from the Weak Tower Alignment Conjecture in that the towers are immediately aligned with the geodesics. However, it is possible that this conjecture is too strong. For instance it implies that there exist heavy towers as one backtracks along the geodesic, and these towers might not be BPS. Nevertheless, it is still possible that the Tower Alignment Conjecture still holds, and in this paper I find examples where it does hold.

The Tower Alignment Conjecture does not imply the Tower SWGC, but it implies the Weak Tower Alignment Conjecture and the Sharpened Distance Conjecture. This is because the $\alpha_\parallel$ condition, following the discussion around equation \eqref{e.tensionless}, implies that there are light towers as one goes far enough down the geodesic. To obtain the Tower SWGC, there needs to be an additional ingredient.

In order to bridge the gap and obtain the Tower SWGC, I propose the following conjecture about the dense nature of infinite-distance geodesics:
\begin{namedconjecture}[Dense Direction Conjecture]
	Consider an arbitrary point $\phi$ in the moduli space, and all geodesics passing through $\phi$ that go to infinite distance limits. The set of directions of these geodesics at $\phi$ is dense in the set of all directions in the tangent space at $\phi $.
\end{namedconjecture}
\noindent This conjecture concerns only the geodesics of the moduli space, and it is independent of the particle and string spectra.

Remarkably, when the Dense Direction and the Tower Alignment Conjectures are simultaneously true, then the Tower SWGC and the Sharpened Distance Conjecture both follow as consequences! This is a major result of this paper, and the later sections are devoted to shining light on examples where this occurs.

There are other related conjectures that frequently arise in the examples considered in this paper.

First, the examples studied in this paper motivate a conjecture about heavy towers that accompany infinite-distance limits:
\begin{namedconjecture}[Heavy Towers Conjecture]
	Fixing a point $\phi_1$ in the moduli space, the theory at a sufficiently far away point $\phi_2$ in the moduli space has an infinite tower of heavy particles, with characteristic mass scaling as
	\begin{align}
		m\sim e^{|\alpha|||\phi_1,\phi_2||},\qquad \text{as}\qquad ||\phi_1,\phi_2||\rightarrow\infty,
	\end{align}
	where $||\phi_1,\phi_2||$ is the length of the shortest geodesic in $\mathcal M$ connecting $\phi_1$ and $\phi_2$, and $|\alpha|\geq 1/\sqrt{d-2}$.
\end{namedconjecture}
\noindent This conjecture implies that under any infinite distance limit, there exists a tower that becomes exponentially heavy. This conjecture holds when the non-sliding Tower SWGC holds (see Section \ref{s.sometimes}), because in that case the convex hull nature of the non-sliding Tower SWGC implies a differential equation for towers whose $\alpha$-vectors are anti-aligned with infinite distance limits. Remarkably, the Heavy Towers Conjecture seems to hold for the cases considered in this paper where $\alpha$-vectors do slide, up to issues with oscillators of high-tension strings which will be the subject of future work.

It is also interesting to strengthen the Weak Tower Alignment Conjecture, but not all the way to the Tower Alignment Conjecture, by specifying the locations on the geodesics when the towers align in terms of the moduli-dependent species scale of \cite{vandeHeisteeg:2023ubh} and the masses of the towers: 
\begin{namedconjecture}[Light Tower Alignment Conjecture]
	Consider an arbitrary point $\phi$ in the moduli space and an arbitrary infinite distance geodesic $\gamma$ passing through $\phi$. Then there exists a tower with an $\alpha$-vector such that its projection along the geodesic satisfies
	\begin{align}
\alpha_\parallel\geq 1/\sqrt{d-2},
	\end{align}
	provided that the aligned towers are lighter than the species scale.
\end{namedconjecture}
\noindent This conjecture is stronger than the Weak Tower Alignment Conjecture but weaker than the Tower Alignment Conjecture. Additionally, it connects the Tower Alignment with the species scale, and possibly the Desert Points \cite{Long:2021jlv,vandeHeisteeg:2022btw}. It is also possible that the conjectures of \cite{Rudelius:2023mjy} could be connected with this conjecture. This is the subject of future work.

On the other hand, one can also conjecture a statement even stronger than the Tower Alignment Conjecture:
\begin{namedconjecture}[Strong Tower Alignment Conjecture]
	For any pair of points $\phi_1$ and $\phi_2$, and an infinite distance geodesic $\gamma_1$ passing through $\phi_1$, there exists a geodesic $\gamma_2$ passing through $\phi_2$ going to the same infinite distance limit as $\gamma_1$ such that there is a single tower satisfying $\alpha_\parallel\geq 1/\sqrt{d-2}$ everywhere along both $\gamma_1$ and $\gamma_2$.
\end{namedconjecture}
\noindent While this conjecture holds in 9d and 10d maximal supergravity, it is not yet known to what extent this conjecture holds in the moduli space of 16 supercharge theories. While the Dense Direction Conjecture and Tower Alignment Conjecture imply the Tower SWGC and the Sharpened Distance Conjecture, it is possible that the Strong Tower Alignment Conjecture implies stronger statements.

In the maximal and half-maximal supergravity examples, I also find the following conjecture holds.
\begin{namedconjecture}[Gradient Flow Convergence Conjecture]
	The gradient flows of the logarithms of masses of towers converge to geodesics.
\end{namedconjecture}
\noindent In maximal supergravity examples, the gradient flows immediately align with geodesics, as will be demonstrated. Also, in half-maximal supergravity examples studied in this paper, the $\alpha$-vectors restricted on the dilaton-radion plane do not immediately give geodesics, though they eventually tend to geodesics.

It is possible that the $\alpha$-vectors of some subset of particles or branes, perhaps the BPS particles and branes, always point immediately along geodesics. If possible, this might allow one to use the spectrum to distinguish infinite-distance geodesics from other geodesics that do not go to infinite distance limits.

\section{IIB in 10d \label{s.IIB10d} }
In this section, I examine how the above conjectures hold in 10d IIB string theory.

In IIB string theory in 10d, the moduli space is non-flat and the particles come from $(p,q)$-string oscillators. The $(p,q)$-strings have $\alpha$-vectors of length $1/\sqrt{d-2}$ each. However, the $\alpha$-vectors of these $(p,q)$-strings are moduli-dependent and just barely manage to align everywhere with non-straight infinite-distance geodesics. This provides a highly nontrivial testing ground for the conjectures in Section \ref{s.conjectures}.

In this example, the Dense Directions Conjecture holds, and for each infinite distance geodesic there is a $(p,q)$-string whose $\alpha$-vector aligns everywhere with the geodesic. This implies the Stringy SWGC and the Sharpened Distance Conjecture. If one also assumes that the $(p,q)$-strings have oscillator towers with masses that scale with the square roots of the tensions, then the Tower Alignment Conjecture, Tower SWGC, and Heavy Tower Conjecture hold as well.

\subsection{Geodesics}
\begin{figure}
\begin{center}
\includegraphics[width = 80mm]{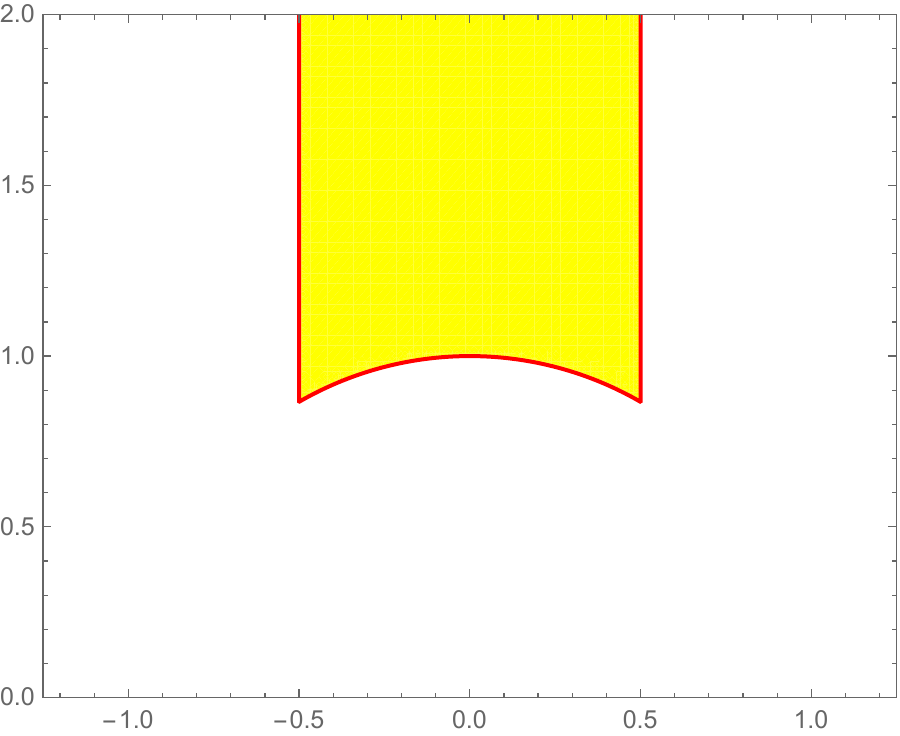}
\end{center}
\caption{Fundamental domain of IIB string theory in 10d.}
\label{f.keyhole}
\end{figure}

In IIB string theory, the moduli space is the upper half-plane quotiented by the action of SL$(2,\mathbb Z)$. We can restrict our attention to the ``keyhole region" fundamental-domain, which is
\begin{align}
	\mathcal M_\text{10d IIB}=\left\{z=x+iy\ |\ x\in[-1/2,1/2], x^2+y^2\geq 1\right\}.
\end{align}
This region is depicted in Figure \ref{f.keyhole}.

As can be seen from the effective action (stated later in equation \eqref{e.10deff}) the metric on this space is that of the Poincar\'e half plane,
\begin{align}
	ds^2=\frac{dx^2+dy^2}{y^2}.
\end{align}
Using the definition of infinite distance limits from Section \ref{s.conjectures}, this metric allows for only one unique infinite distance limit in the fundamental domain. This limit occurs with $y\rightarrow \infty $.

To see whether the conjectures from Section \ref{s.conjectures} hold in this case, we need to first obtain all of the geodesics that go to $y\rightarrow \infty$. This at first seems complicated because, in general, geodesics can pass through an edge of the fundamental domain, reappear on another edge (due to the identifications of edges), and have this process repeat several times until eventually the geodesic is a vertical geodesic that goes to $y\rightarrow \infty$.

\begin{figure}
\begin{center}
\includegraphics[width = 80mm]{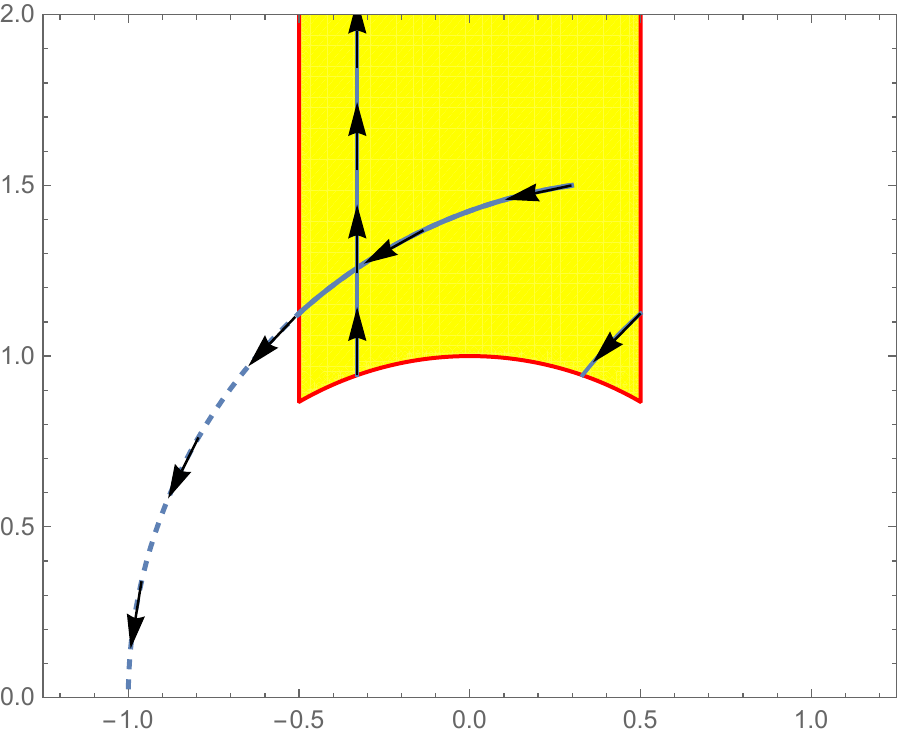}
\end{center}
\caption{Example of an infinite-distance geodesic that starts at $\tau=0.3+1.5i$ and goes to the point $-1$ on the upper-half-plane covering space.}
\label{f.keyholegeodesic}
\end{figure}

These infinite distance geodesics are easier to analyze in the Poincar\'e upper-half plane covering space than in the fundamental domain. To proceed, there are two facts to note.
\begin{enumerate}
	\item The geodesics on the Poincar\'e half-plane are circles whose centers are on the real axis. This includes vertical lines, as these can be thought of circles whose centers are on the real axis at $x=\pm \infty$.
	\item The rational points $Q=m/n$ on the $x$-axis are the points on the Poincar\'e half-plane that are identified with $y\rightarrow \infty$ in the fundamental domain.
\end{enumerate}
For an arbitrary point $z=x+iy$ in the fundamental domain, there is a one-to-one mapping between infinite-distance geodesics in the fundamental domain passing through $z$ and circles in the Poincar\'e half-plane that are centered on the $x$-axis, pass through $z$, and pass through rational points (or $x=\pm\infty$) on the $x$-axis. In particular, for any infinite-distance limit geodesic in the fundamental domain, until this geodesic crosses one of the boundaries of the fundamental domain, it coincides with a circle in the Poincar\'e half-plane that ends on a rational point. This is demonstrated in Figure \ref{f.keyholegeodesic}.

In practice, these ``rational circles" in the Poincar\'e half-plane are easier to deal with than the geodesics in the fundamental domain, so we will perform our analysis in the upper-half plane, instead of in the fundamental domain.

Suppose we start at a point $\tau=\tau_1+i\tau_2$ in the fundamental domain and we are interested in the geodesic circle that passes through both $\tau$ and the rational point $Q$ on the $x$-axis. The points $z=x+iy$ on the circle satisfy
\begin{align}
	(x-c)^2+y^2=r^2.
\end{align}
So, if the circle is to pass through the point $\tau$, then
\begin{align}
	(\tau_1-c)^2+\tau_2^2=r^2,
\end{align}
and if the circle is to also pass through the rational point $Q$ on the $x$-axis, then
\begin{align}
	(Q-c)^2=r^2.
\end{align}
For the circle to pass through $\tau$ and $Q$, then the center and radius of this circle must be
\begin{align}
	c= \frac{\tau_1^2+\tau_2^2-Q^2}{2 \tau_1-2 Q},\qquad r= \frac{(Q-\tau_1)^2+\tau_2^2}{2 | Q-\tau_1| }.
\end{align}
All infinite-distance geodesics are described by circles of this type.

\begin{figure}
\begin{center}
\includegraphics[width = 80mm]{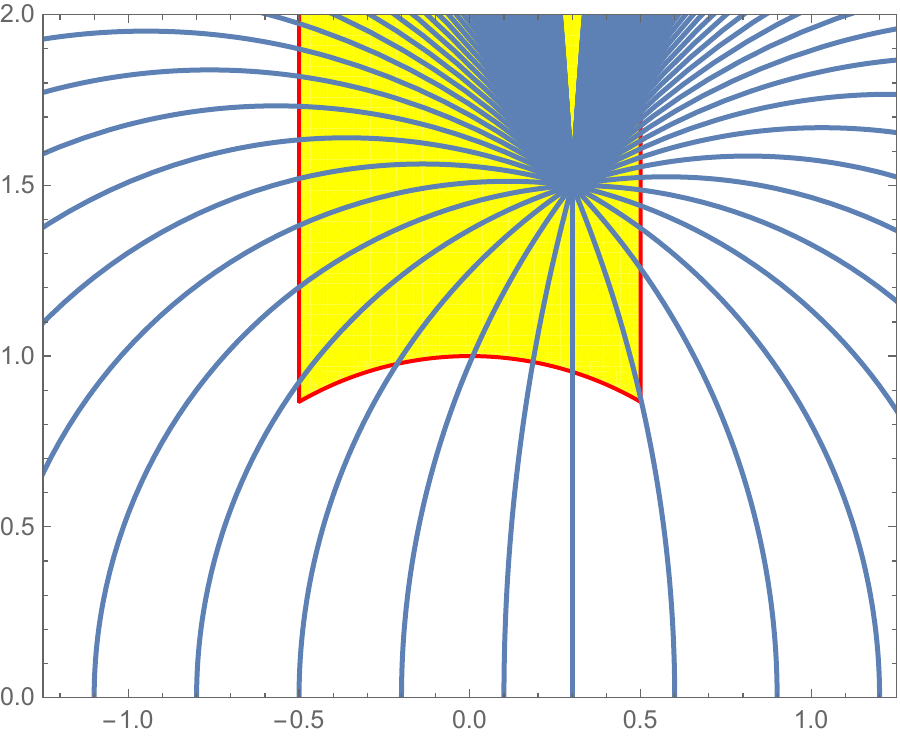}
\end{center}
\caption{Denseness of geodesics. These are some infinite distance geodesics that pass through $\tau=.3+1.5i$ and go to infinite distance limits (i.e., these geodesics end on rational points on the $x$-axis). These geodesics pass through the point $\tau$ in a set of directions that is dense in the set of all directions in the tangent space at $\tau$.}
\label{f.dense}
\end{figure}

The Dense Direction Conjecture is satisfied. To see this, consider an arbitrary point $\tau$ in the fundamental domain. Consider the set of all geodesic circles that pass through $\tau$ and go through rational points on the $x$-axis. This set of circles passes through the point $\tau$ in a set of directions that is dense in the set of all directions in the tangent space at $\tau$. See Figure \ref{f.dense}.

\subsection{$\alpha$-vectors of $(p,q)$-strings}
The $(p,q)$-string has tension, using for instance (14.1.9) of \cite{Polchinski:1998rr},
\begin{align}
	\tau_{(p,q)}&=l_0^{-2}\sqrt{e^\Phi (p+C_0q)^2+e^{-\Phi}q^2}=l_0^{-2}\frac{|p+\tau q|}{\sqrt{\tau_2}}
\end{align}
where
\begin{align}
	\tau=C_0+ie^{-\Phi}.
\end{align}
Let us assume\footnote{As was discussed in Section \ref{s.introduction}, in this section we are technically verifying only the Stringy SWGC as a proxy for the Tower SWGC. The Stringy SWGC does not need this assumption about the oscillators scaling with the square root of the tension.} that the oscillators of $(p,q)$-strings have masses that scale with the square root of the tension,
\begin{align}
	m_{(p,q)}\sim \sqrt{\tau_{(p,q)}}.
\end{align}

To compute the $\alpha$-vectors, we canonically normalize the moduli. This can be obtained by studying the effective action in Einstein frame, which is
\begin{align}
	S_\text{IIB}=\frac{1}{2\kappa_{10}^2}\int d^{10}x\sqrt{-G}\left(R-\frac{\partial \bar \tau \partial \tau}{2\tau_2^2}\right ).\label{e.10deff}
\end{align}
The canonically normalized moduli are, in appropriate Planck units, thus
\begin{align}
	\hat \tau=\frac{\tau_1+i\tau_2}{\sqrt 2\left\langle \tau_2\right \rangle }.
\end{align}
With respect to this modulus, then the $\alpha$-vectors are
\begin{align}
	\vec \alpha_{(p,q)}=(
		-\partial_{\hat\tau_1} \log m_{(p,q)},
		-\partial_{\hat\tau_2} \log m_{(p,q)}).
\end{align}
Explicitly, the $\alpha$-vectors, with respect to canonically normalized moduli, are
\begin{align}
	\vec\alpha =(\alpha_{\hat \tau_1},\alpha_{\hat \tau_2}),\qquad
	\alpha_{\hat \tau_1}=\frac{q\tau_2(p+\tau_1q)}{\sqrt 2|p+\tau q|^2},\qquad
	\alpha_{\hat \tau_2}=\frac{q^2\tau_2^2-(p+\tau_1q)^2}{\sqrt 8|p+\tau q|^2},\label{e.alphaIIB10d}
\end{align}
and these all have length
\begin{align}
	|\vec \alpha|=\frac 1{\sqrt{8}}=\frac 1{\sqrt{d-2}}.
\end{align}

\begin{figure}
\begin{center}
\includegraphics[width = 80mm]{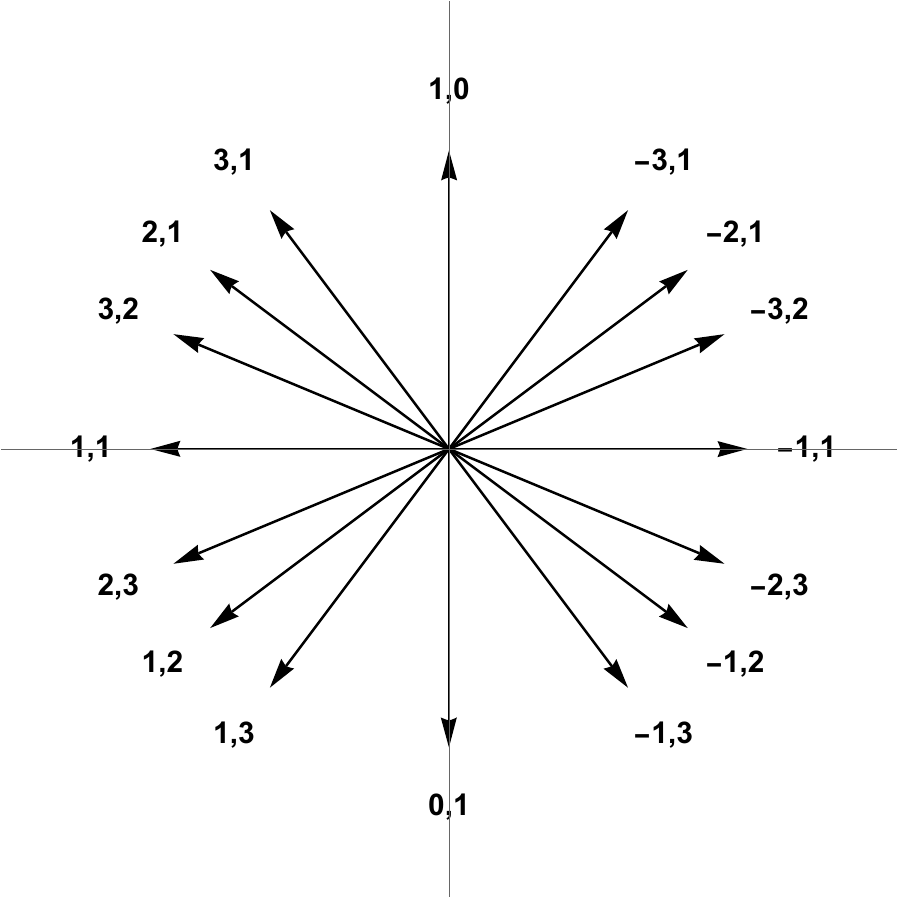}
\end{center}
\caption{$\alpha$-vectors for some $(p,q)$-strings at $\tau=i$. If all $(p,q)$-string $\alpha$-vectors had been included, they would densely populate the circle of radius $1/\sqrt{7}$.}
\label{f.alpharay}
\end{figure}

\begin{figure}
\begin{center}
\includegraphics[width = 80mm]{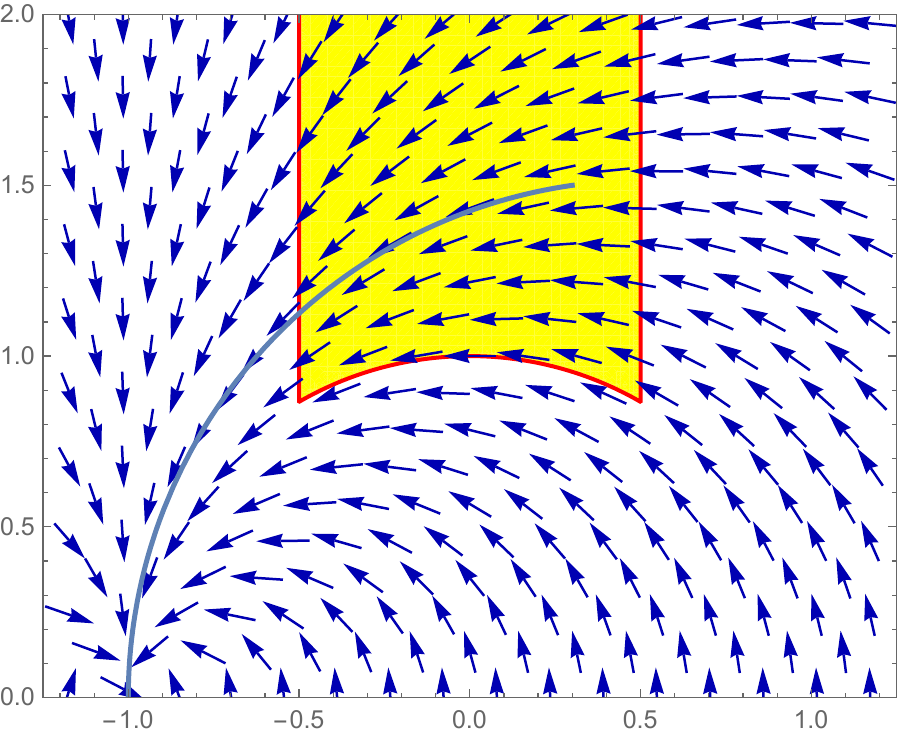}
\end{center}
\caption{$\alpha$-vector field for $(1,1)$ strings on the entire Poincar\'e upper-half plane. The blue line is an infinite-distance geodesic going to the point $-1$ on the $x$-axis. Note that this infinite distance geodesic is an integral curve of the $\alpha$-vector field.}
\label{f.alphafield}
\end{figure}

At each point $\tau$ in moduli space, there exists countably infinite set of $\alpha$-vectors in the tangent space $T_\tau(\mathcal M)$ at $\tau$. These vectors densely populate a circle of radius $1/\sqrt{d-2}$ (see Figure \ref{f.alpharay}). Also, each $(p,q)$-string has an associated $\alpha$-vector field (see Figure \ref{f.alphafield}).

From Figure \ref{f.alphafield}, one can see a remarkable one-to-one correspondence between infinite-distance geodesics and $\alpha$-vectors! First, for every infinite-distance geodesic going through the point $\tau$, there is single $(p,q)$-string whose $\alpha$-vector is aligned with that geodesic. Second, $(p,q)$-string $\alpha$-vector field generates a vector flow, and any such flow is an infinite-distance geodesic (when extended over all of the Poincar\'e half plane) that goes to the point $-p/q$ on the $x$-axis.

\subsection{Testing the conjectures}
Consider an arbitrary geodesic going through the rational point $Q$ on the $x$-axis and the arbitrary point $z=x+iy$ in the upper-half-plane. A $(p,q)$-string $\alpha$ is perpendicular to this circle if and only if
\begin{align}
	0=\vec \alpha \cdot ((x-c)\unit x+y\unit y)\propto \frac pq+Q.
\end{align}
So, the $(p,q)$-string has an $\alpha$-vector whose flow generates translation to the infinite distance point $-p/q$ on the real axis of the Poincar\'e half-plane.

\begin{figure}
\begin{center}
\includegraphics[width = 80mm]{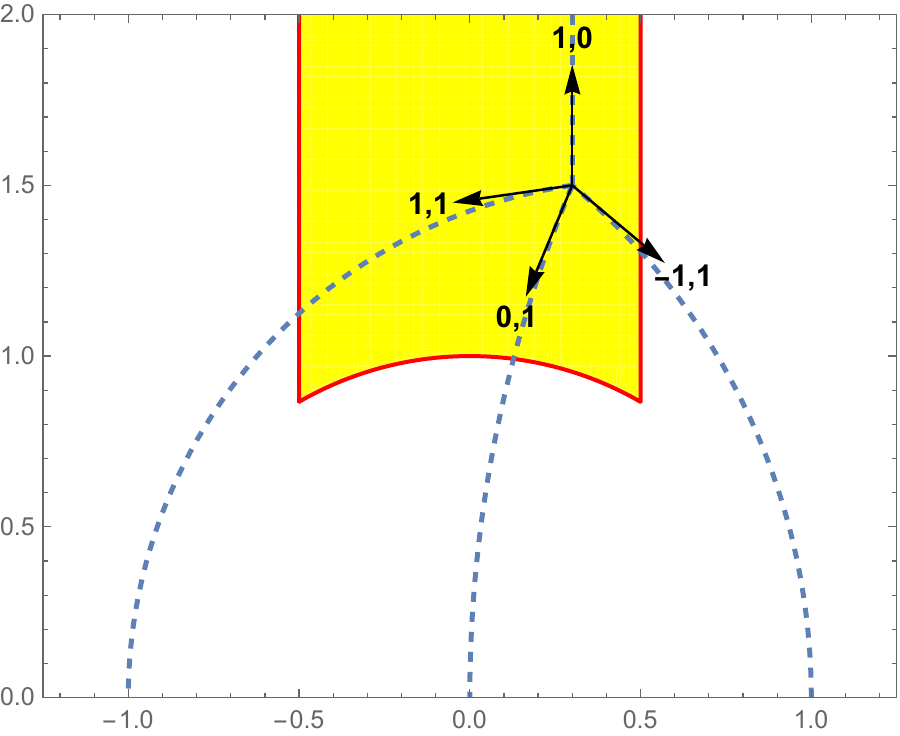}
\end{center}
\caption{$\alpha$-vectors for $(1,1)$, $(-1,1)$, $(0,1)$ and $(1,0)$ strings are parallel to geodesics going to $-1$, $1$, $0$, and $i\infty$.}
\label{f.alpharaygeodesics}
\end{figure}

For each $(p,q)$-string, the $\alpha$-vector field is parallel to every infinite-distance geodesic that goes to the rational point $-p/q$ on the $x$-axis. See Figure \ref{f.alpharaygeodesics}.

The Strong Tower Alignment Conjecture also holds. The $\alpha$-vector field of the $(p,q)$-string oscillators is always parallel to any geodesic that goes to the rational point $-p/q$, and the projections of the $(p,q)$-string $\alpha$-vectors along any such geodesic is precisely $\alpha_\parallel=1/\sqrt{d-2}$, thus satisfying the Strong Tower Alignment Conjecture.

There are several other interesting things to note. In a particular infinite distance limit, there is precisely only one emergent string at a time. Also, in these infinite distance limits, all of the other strings acquire large tensions. Thus, their oscillations become the heavy towers of the Heavy Towers Conjecture.

While there are infinitely many infinite distance geodesics passing through each point in moduli space, the directions of these geodesics on the tangent plane are actually a set of measure zero. This is because the set of directions at each point include all angles between $0$ and $2\pi$, which is an uncountable set, but the geodesics to infinite distance limits are a countable subset, and thus they have measure zero.

\section{32 supercharges in 9d \label{s.329d} }
In this section, I explicitly demonstrate that the conjectures of section \ref{s.conjectures} hold for IIB string theory on a circle, which is equivalent to M-theory on a two-torus and also to IIA string theory on a circle. This demonstration does not rely on the Stringy SWGC as a proxy for the Tower SWGC because in this case the 1/2 BPS particles generate the Tower SWGC.

\subsection{Geodesics}
In $d=9$ dimensional maximal supergravity, there are three moduli, all originating from the eleven-dimensional graviton coming from M-theory on $T^2$. To determine their couplings, we reduce the $D=11$ dimensional Einstein-Hilbert action
\begin{align}
	S=\frac 1{2\kappa_{11}^2}\int d^{11}x\sqrt{-g_{(11)}}\mathcal R[g_{(11)}],
\end{align}
with the ansatz
\begin{align}
	ds^2_D= \Vert g \Vert^{-\frac{1}{d-2}} g_{\mu \nu}dx^\mu dx^\nu+g_{m n} d y^m d y^n,\label{eq:metricansatz}
\end{align}
where $m,n$ index the $D-d=2$ compact directions, $\mu, \nu$ index the $d$ noncompact directions, $y^m \cong y^m + 2\pi R$, and $\Vert g \Vert = \det g_{mn}$. It is convenient to decompose $g_{mn}$ in terms of volume and shape parameters $U$ and $\tau = \tau_1 + i \tau_2$, where
\begin{align}
g_{m n} = e^U \frac 1{\tau_2}\begin{pmatrix}1&\tau_1\\\tau_1&|\tau|^2\end{pmatrix}.
\end{align}
Here, the moduli space is a product manifold, where the the modulus $U$ controls the volume of the two-torus, and the shape parameters $\tau=\tau_1+i\tau_2$ control the shape of the two torus and are the same moduli axio-dilaton moduli from IIB string theory in 10d. With this, the Einstein-moduli sector of the dimensionally reduced action is
\begin{align}
S_9=\frac 1{2\kappa_9^2}\int d^9x\sqrt{-g}\left(\mathcal R-\frac 9{14}(\partial U)^2-\frac{(\partial \tau_1)^2+(\partial \tau_2)^2}{2\tau_2^2}\right).
\end{align}

The moduli space for $U$ is flat, whereas the moduli space for $\tau$ has the same Poincar\'e half-plane metric as in the 10d IIB string theory case.

All of the infinite distance geodesics can be described by the three following classes:
\begin{enumerate}
	\item $U\rightarrow\pm \infty$.
	\item The imaginary part of $\tau$ approaches $i\infty$, or $\tau$ approaches a rational point on the $x$-axis of the half-plane covering space.
	\item A combination of the two above limits.
\end{enumerate}

All infinite distance geodesics are given by two-parameters $\theta$ and $Q$, where $\theta\in[0,\pi]$ and $Q$ is a rational number. The geodesics of unit speed are parametrized by $t$ and given by
\begin{align}
	\gamma_{\theta,Q}(t)=(\cos \theta\  t,\sin \theta\  \gamma_Q(t)),\label{e.9dgeofam}
\end{align}
where $\gamma_Q(t)$ is the unit-speed geodesic discussed in the 10d IIB case that ends on rational points $Q$ on the $x$-axis of the Poincar\'e half-plane.

\subsection{$\alpha$-vectors}
Working in modified nine-dimensional Planck units where $2 \kappa_9^2 = (2\pi)^6$ for convenience, the 1/4 BPS particles have masses
\begin{align}
m_{p,q,w}=\frac{|p+\tau q|}{\sqrt{\tau_2} R}e^{-\frac 9{14} U}+R^{4/3} |w| e^{\frac 67U},
\end{align}
where $p, q\in \mathbb{Z}$ are the Kaluza-Klein charges and $w \in \mathbb{Z}$ is the M2 brane winding charge. These particles are $1/2$ BPS when either $w=0$ or $p=q=0$.

At a particular point in moduli space, the canonically normalized moduli are
\begin{align}
\hat \phi^a=(\hat U,\hat \tau_1,\hat \tau_2)=\Biggl(\sqrt{\frac{9}{14}}U,\frac{\tau_1}{\sqrt{2} \langle \tau_2\rangle},\frac{\tau_2}{\sqrt{2} \langle \tau_2\rangle}\Biggr),
\end{align}
where $\langle \tau_2\rangle$ is the vacuum expectation value of $\tau_2$ at the point in question. In this orthonormalized bases, the $\alpha$-vectors for the $1/4$ BPS particles are
\begin{subequations}
\begin{align}
\alpha_{\hat{U}} &= -\frac{4  \sqrt{\tau_2} e^{\frac{3 U}{2}} R^{7/3} |w| -3  |p+\tau q|}{\sqrt{14} \left(|p+\tau q|+ \sqrt{\tau_2} e^{\frac{3 U}{2}} R^{7/3}  |w|\right)},\\
\alpha_{\hat{\tau}_1} &=- \frac{\sqrt{2} q \tau_2 (p+\tau_1 q)}{|p+\tau q| \left(|p+\tau q|+\sqrt{\tau_2} e^{\frac{3 U}{2}} R^{7/3}  |w| \right)},\\
\alpha_{\hat{\tau}_2} &= -\frac{q^2 \tau_2^2-(p+\tau_1 q)^2}{\sqrt{2} |p+\tau q| \left(|p+\tau q|+ \sqrt{\tau_2} e^{\frac{3 U}{2}} R^{7/3}  |w|  \right)}.
\end{align}
\end{subequations}

\begin{figure}
\begin{center}
\center
\includegraphics[width=3.5in]{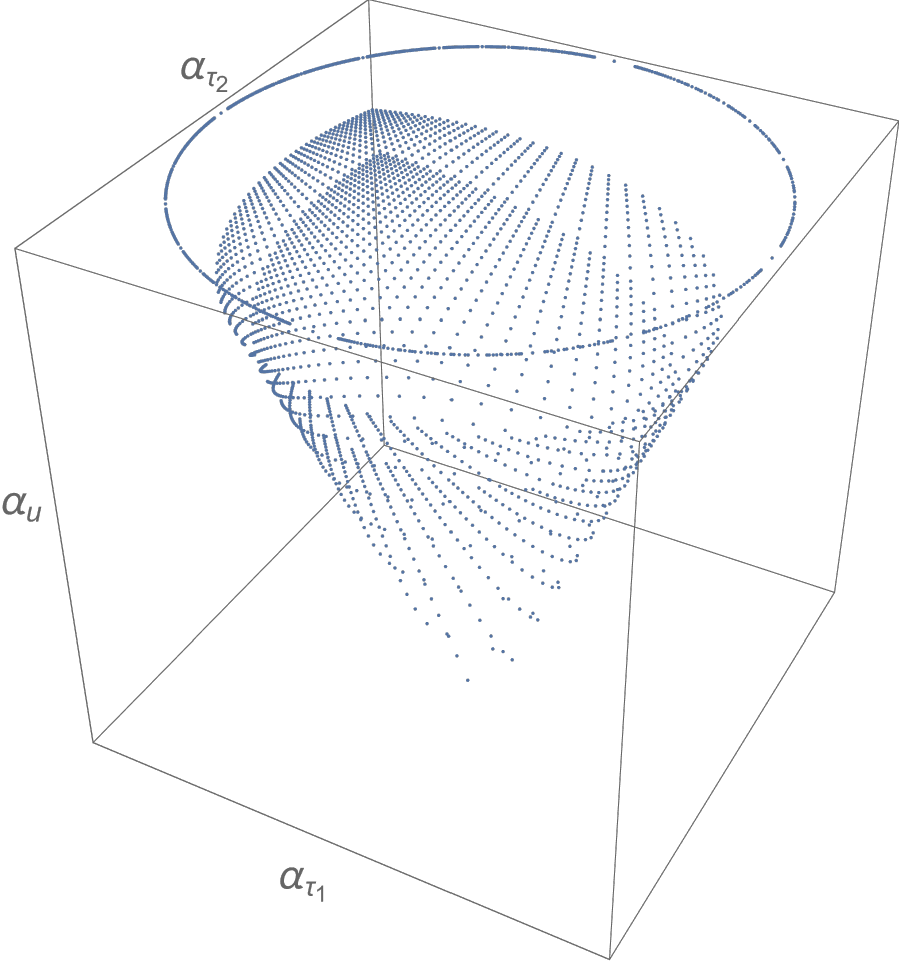}
\caption{
Some 1/4 BPS $\alpha$-vectors from M-theory on $T^2$. Here, the 1/4 BPS vectors $\vec\alpha$ run over the values $w\in\{0,1,2\}$, and $p,q\in\{-25,\dots,25\}$.
\label{f.alphaconedots}}
\end{center}
\end{figure}

\begin{figure}
\begin{center}
\includegraphics[width = 80mm]{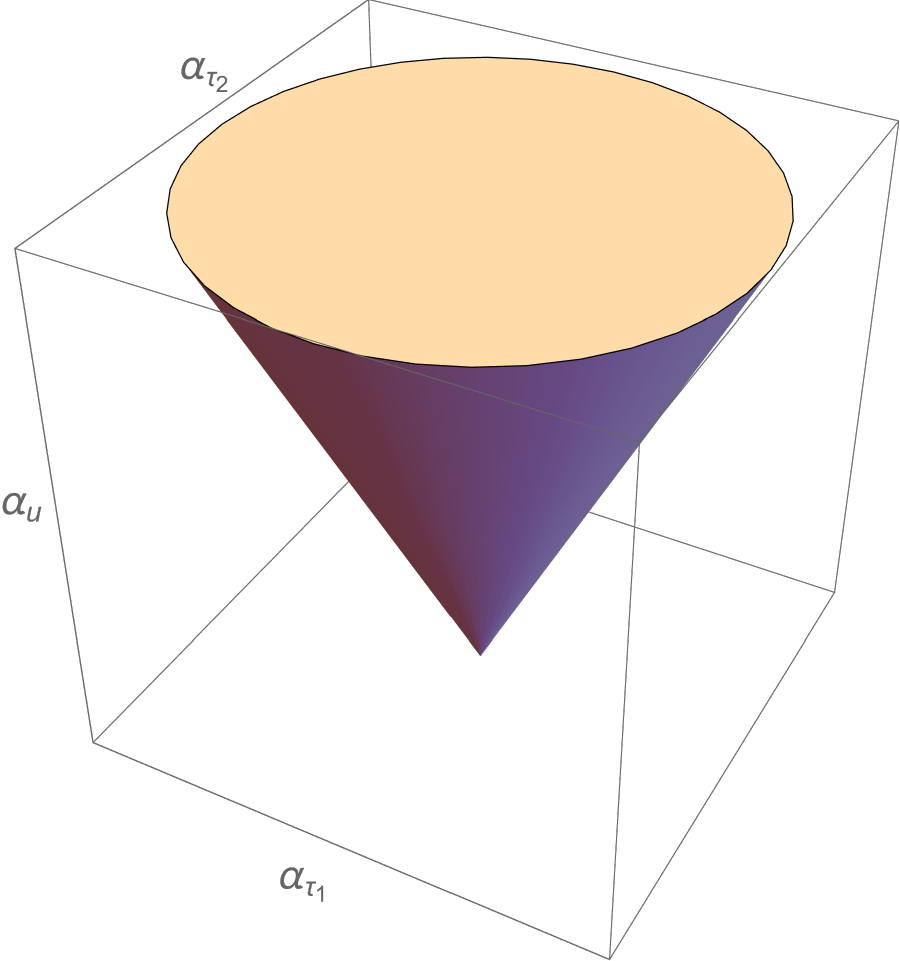}
\end{center}
\caption{Closure of the convex hull of $\alpha$-vectors of M-theory on $T^2$.}
\label{f.alphacone}
\end{figure}

As illustrated in Figures \ref{f.alphaconedots} and \ref{f.alphacone}, these $\alpha$-vectors densely populate a cone. At the tip of the cone lie the 1/2 BPS states with $p=q=0$ and $w\neq 0$, corresponding to an M2 brane wrapped $w$ times on $T^2$:
\be
\vec \alpha_9^\text{winding}=\left(-\sqrt{\frac87},0,0\right).\label{e.alphawinding9d}
\ee
 The base of the cone is populated by the 1/2 BPS Kaluza-Klein modes with $w=0$ but nonzero $p$ or $q$:
 \be
\vec\alpha_9^\text{KK}\in\left\{\left(\frac {3}{\sqrt {14}},\alpha_{\tau_1},\alpha_{\tau_2}\right)\in \mathbb R^3:\alpha_{\tau_1}^2+\alpha_{\tau_2}^2=\frac12\right\}.\label{e.alphaKK9d}
\ee
 The remaining 1/4 BPS states lie along the cone somewhere between its tip and circular base. From this example, we see that the convex hull generated by the 1/2 BPS states contains the convex hull generated by the 1/4 BPS states.

The purely winding $\vec \alpha_9^\text{winding}$-vectors are a distance $\sqrt{8/7}$ from the origin, and so too are the Kaluza-Klein $\vec \alpha_9^\text{KK}$-vectors. Thus, the points on the cone closest to the origin lie on a circle halfway between the tip of the cone and the circular base of the cone.

\subsection{Testing the conjectures}
The projection $\alpha_\parallel$ of an $\alpha$-vector along an infinite distance geodesic $\gamma_{\theta,Q}(t)$ is given by
\begin{align}
	\alpha_\parallel=\vec \alpha \cdot \dot \gamma_{\theta,Q}(t).
\end{align}

For fully wrapped M2-branes with no KK-momenta, the $\alpha$-vectors are given by \eqref{e.alphawinding9d}, and the projection of such vectors along an infinite distance geodesic $\gamma_{\theta,Q}(t)$ is given by
\begin{align}
	\alpha_\parallel^\text{winding}=\vec \alpha^\text{winding} \cdot \dot \gamma_{\theta,Q}(t)=-\sqrt{\frac{8}{7}}\cos \theta 
\end{align}

For $(p,q)$-KK modes with no winding charge, the $\alpha$-vectors are given by \eqref{e.alphaKK9d}, which is
\begin{subequations}
\begin{align}
\alpha_{\hat{U}}^\text{KK} &= \frac{ 3 }{\sqrt{14} },\\
\alpha_{\hat{\tau}_1}^\text{KK} &=- \frac{\sqrt{2} q \tau_2 (p+\tau_1 q)}{|p+\tau q|^2},\\
\alpha_{\hat{\tau}_2}^\text{KK} &= -\frac{q^2 \tau_2^2-(p+\tau_1 q)^2}{\sqrt{2} |p+\tau q|^2}.
\end{align}
\end{subequations}
It is convenient to relate the $\tau$ components of these $\alpha$-vectors to the 10d IIB $\alpha$-vectors for $(p,q)$-strings given by equation \eqref{e.alphaIIB10d}. This results in
\begin{align}
	(\alpha_{\hat{\tau}_1}^\text{KK},\alpha_{\hat{\tau}_1}^\text{KK})_{\text{II}}^\text{9d} &=2\vec\alpha_\text{IIB}^\text{10d}.
\end{align}
Thus, the 9d $(p,q)$-KK mode is given by
\begin{align}
	\vec\alpha_{p,q}^\text{9d KK}=\left(\frac 3{\sqrt{14}},2\vec\alpha_{p,q}^\text{9d KK}\right).
\end{align}
For each geodesic $\gamma_{\theta,Q}(t)$, it suffices to consider the $(p,q)$ KK modes such that $p/q=\pm Q$. The projection of the $\alpha$-vectors for these modes satisfy
\begin{align}
	\alpha_\parallel^\text{KK}=\vec \alpha_{p,q}^\text{9d KK}\cdot \dot \gamma_{\theta,\pm p/q}(t)=\frac{3}{\sqrt{14}}\cos \theta \mp   \frac 1{\sqrt 2}\sin \theta .
\end{align}

These projections allow us to test the conjectures of Section \ref{s.conjectures}. There are three different regimes of $\theta\in[0,\pi]$ to consider in this test. For the different regimes of $\theta$, the following conditions hold everywhere along the geodesics:
\begin{enumerate}
	\item $\boldsymbol{\theta \leq \arctan \sqrt 7}$: In this case, the $(p,q)$ KK modes with $-p/q=Q$ have $\alpha_\parallel^\text{KK} >1/\sqrt 7$, and the winding modes have $\alpha_\parallel^\text{winding}\leq -1/\sqrt 7$, thus satisfying the Tower Alignment and Heavy Tower Conjectures.
	\item $\boldsymbol{\arctan \sqrt 7\leq \theta \leq \pi- \arctan \sqrt 7}$: In this case, the $(p,q)$ KK modes with $-p/q=Q$ have $\alpha_\parallel^\text{KK} \geq 1/\sqrt 7$, and $(p,q)$ KK modes with $p/q=Q$ have $\alpha_\parallel^\text{KK} \leq- 1/\sqrt 7$, thus satisfying the Tower Alignment and Heavy Tower Conjectures.
	\item $\boldsymbol{  \pi-\arctan \sqrt 7\leq \theta }$: In this case, winding modes have  $\alpha_\parallel^\text{winding} \geq 1/\sqrt 7$ and the $(p,q)$ KK modes with $p/q=Q$ have $\alpha_\parallel^\text{KK} <- 1/\sqrt 7$, thus satisfying the Tower Alignment and Heavy Tower Conjectures.
\end{enumerate}
Thus, along all of these geodesics, the Tower Alignment and Heavy Tower Conjectures hold from the 1/2 BPS states!

\section{16 supercharges in 9d \label{s.169d}}

The examples considered above have maximal supergravity, and so it is interesting to see to what extent the proposals of Section \ref{s.conjectures} apply to half-maximal supergravity cases. While it is possible that all of the proposals apply to half-maximal supergravity cases, testing this fully is beyond the scope of this paper and left to future work. Nevertheless, in this section we investigate some tests of the conjectures of Section \ref{s.conjectures}.

The maximal supergravity cases have some major differences from the half-maximal supergravity cases. In the maximal supergravity cases studied above, the Stringy SWGC is satisfied by BPS states, and in 9d and in lower dimensions the Tower SWGC is satisfied by BPS particles. Meanwhile, in 9d theories with 16 supercharges, the Tower SWGC (and also Stringy SWGC) convex hulls are not entirely generated by BPS states \cite{Etheredge:2023odp}. As demonstrated in \cite{Etheredge:2023odp}, the non-BPS states involved in generating the Tower SWGC have $\alpha$-vectors that vary with the location in the moduli space, and as a result the shape of the convex hull of $\alpha$-vectors depends on the location in the moduli space. This is a feature that does not occur in maximal supergravity, where the shape of the convex hull does not change depending on the location in the moduli space. As studied in \cite{Etheredge:2023odp}, this shape deformation in half-maximal supergravity is connected with running decompactifications. Hence, even limited tests of these 9d half-maximal supergravity theories greatly expand the scope of plausibility of the conjectures of Section \ref{s.conjectures}.

In this section, I perform limited tests\footnote{Many of the tests, or similar tests, also occur in \cite{Etheredge:2023odp}, so the reader should consult that reference for more details, discussions, and nontrivial calculations. See also \cite{Polchinski:1995df,Aharony:2007du} for more details. I will quote some important results from \cite{Etheredge:2023odp} and discuss how they agree with the conjectures of Section \ref{s.conjectures}.} of the conjectures from Section \ref{s.conjectures} in asymptotic limits. This is because the non-BPS generators of the Tower SWGC hull are difficult to study when not in asymptotic limits. Additionally, in these 9d half-maximal supergravity cases, the moduli spaces are high-dimensional, and in this section I consider only the submanifold of moduli space involving just the dilaton and the radion of the 9d theory. Understanding the role of the many axions will be the work of a future paper.

In this section, we consider the 9d theories that come from circle-compactifying either $\mathrm{SO}(32)$ or $\mathrm{E_8\times E_8}$ 10d heterotic theories with all of the Wilson lines turned off.

\subsection{Geodesics}

Let us focus on the dilaton-radion, or $\phi-\rho$, submanifold of the moduli space. There are several different kinds of infinite distance limits. As was shown in \cite{Etheredge:2023odp}, the two-dimensional radion-dilaton submanifold of moduli space is a flat plane. As a consequence, any ray in the radion-dilaton plane is a geodesic to an infinite distance limits. This is true for both the SO$(32)$ and $\mathrm{E_8\times E_8}$ cases. Thus, for the dilaton-radion plane, the Dense Direction Conjecture holds.

It is convenient to express the flat dilaton-radion moduli on the $x-y$ plane such that $y\rightarrow -\infty$ with $x$ held constant is the emergent heterotic string limit. In what follows, I refer to $y\rightarrow-\infty$ limits as the ``weak coupling limit" of heterotic string theory. Strong coupling limits, where I or I$'$ string theory are relevant, occur with $y\rightarrow \infty$. These are differentiated by the SO$(32)$ and $\mathrm{E_8\times E_8}$ cases, as discussed in \cite{Etheredge:2023odp}.

In this document, the primary focus is on the strong coupling cases (where $y$ goes to infinity), since in these cases there is running decompactification and the convex hull of $\alpha$-vectors in the Stringy SWGC is limit-dependent, and the conjectures in this case have been already studied in \cite{Etheredge:2023odp}. I demonstrate that the proposals of Section \ref{s.conjectures} pass some tests in the strong coupling regimes. The weak coupling cases also have interesting phenomena, but will be primarily the focus of future research.

\subsection{$\alpha$-vectors}
In this case, I focus on three types of particles: two 1/2 BPS states and the non-BPS KK modes of I$'$ string theory.

In both SO$(32)$ and $\mathrm{E_8\times E_8}$ heterotic string theory in 9d, there are two 1/2 BPS particles. These are the KK modes and the winding modes. For both of these 1/2 BPS particles, the moduli-dependence of the masses of these states remains the same \cite{Etheredge:2023odp}, and thus the $\alpha$-vectors are constant. Their moduli dependence, using the $x$ and $y$ moduli, are given by
	\begin{align}
		m_{\rm w,h}\sim e^{\frac{1}{\sqrt{7}}y+x},\qquad
		m_{\rm KK,h}\sim e^{\frac{1}{\sqrt{7}}y-x}.
	\end{align}

In the strong coupling limits of heterotic string theory, when $y\rightarrow \infty$, there are also non-BPS KK modes from I$'$ string theory. These masses were explicitly computed in \cite{Etheredge:2023odp} for both the $\rm SO(32)$ and $\rm E_8\times E_8$ cases. The masses scale with the $x$ and $y$ moduli as
\begin{subequations}
\begin{align}
	m_{\rm KK,I'}^{\mathrm{SO}(32)}&\sim\frac{(e^{2x}+1)^{3/2}+(e^{2x}-1)^{3/2}}{3e^{4x}+1}e^{\frac{3}{2}x-\frac{5}{2\sqrt{7}}y},\\
	m_{\rm KK,I'}^{\mathrm E_8\times\mathrm  E_8}&\sim e^{-\frac{5}{2\sqrt{7}}y-\frac{1}{2}x}\left(1+e^{-2x}\right)^{-1}.
\end{align}
\end{subequations}

\begin{figure}
\begin{center}
\begin{subfigure}{0.475\textwidth}
\center
\includegraphics[width=70mm]{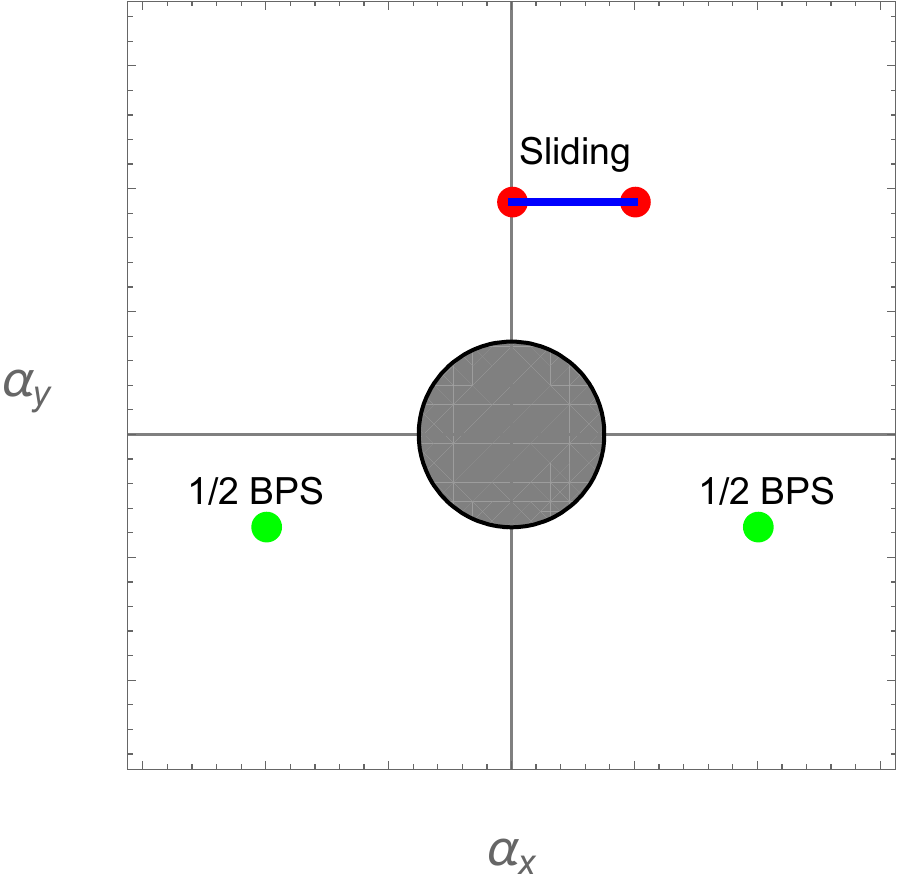}
\caption{$\rm SO(32)$} \label{t.KSslidingSO32}
\end{subfigure}
\begin{subfigure}{0.475\textwidth}
\center
\includegraphics[width=70mm]{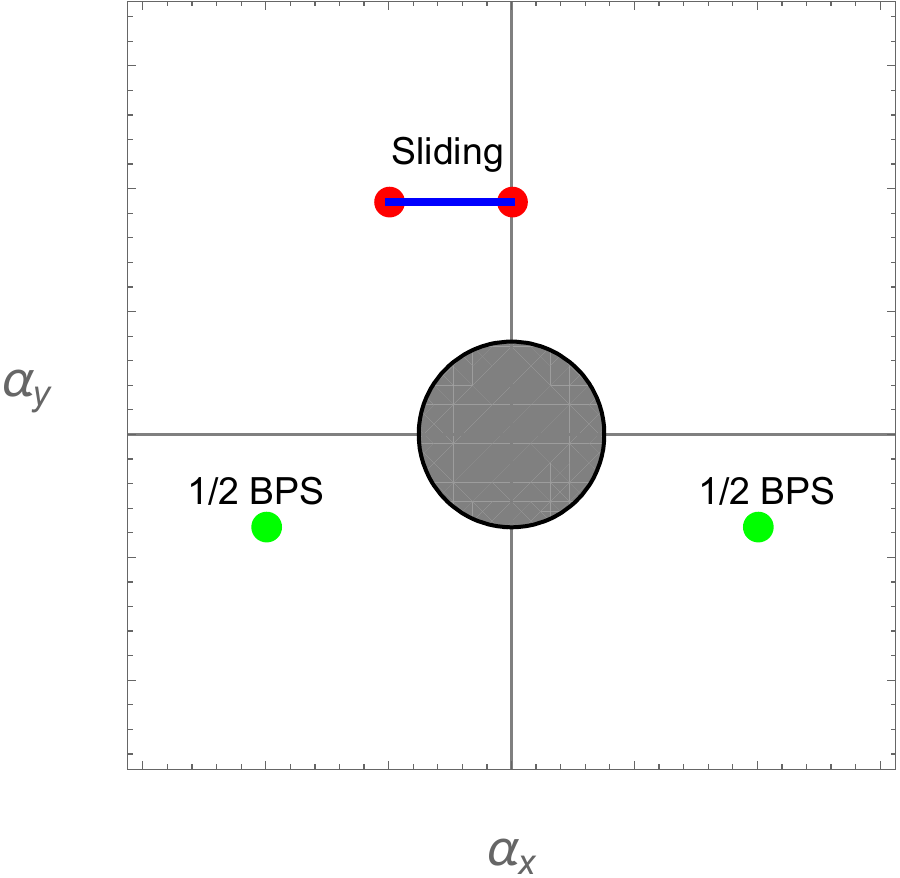}
\caption{$\rm E_8\times E_8$} \label{t.KKSlidingE8E8}
\end{subfigure}
\caption{The sliding of the I$'$ KK modes in the $\rm SO(32)$ and $\rm E_8\times E_8$ cases. The red dots are the I$'$ KK modes that slide, and they slide along the blue line. The green dots are the 1/2 BPS states.\label{f.KKSliding}}
\end{center}
\end{figure}

In these cases, the $\alpha$-vectors in the $x-y$ bases are
\begin{align}
\vec{\alpha}_{\rm w,h}=\left(-1,-\frac{1}{\sqrt{7}}\right),\qquad \vec{\alpha}_{\rm KK,h}=\left(-1,\frac{1}{\sqrt{7}}\right),\notag\\
\vec{\alpha}_{\rm KK,I'}^{\rm SO(32)}=\left(-\frac{3}{2}\left[\frac{2}{\sqrt{1-e^{-4x}}}+1\right]^{-1},\frac{5}{2\sqrt{7}}\right),\\
\vec{\alpha}_{\rm KK,I'}^{\rm E_8\times E_8}=\left(-\frac{1}{2}+\tanh x,\frac{5}{2\sqrt{7}}\right).\notag \label{e.16alphas}
\end{align}
For the $\rm SO(32)$ case, $(\alpha_{\rm KK,I'}^{\rm SO(32)})_x$ is a monotonic function of $x$, with sliding occurring from $\alpha_{x}^{\rm KK,I'}=0$ at $x=0$ to $\alpha_{x}^{\rm KK,I'}=-\frac{1}{2}$ at $x=\infty$. Similarly, for the $\rm E_8\times E_8$ case, there is analogous sliding, but reflected across the $x$-axis. See Figure \ref{f.KKSliding}.

When $x$ is not held constant and $y$ and $x$ simultaneously grow linearly, these I$'$ KK towers slide in such a way so that emergent string limits are not obstructed in the $\mathrm{SO}(32)$ case \cite{Etheredge:2023odp}.

\subsection{Testing the conjectures}

Let us first examine strong-coupling limits. When $x$ is held constant and $y$ goes to $\infty$, then the I$'$ KK modes provide light towers. In such limits the $\alpha$-vectors of the I$'$ KK modes have
\begin{align}
	\alpha_\parallel=\vec \alpha\cdot \unit{y}> 1/\sqrt{d-2}
\end{align}
Meanwhile, if $y$ and $x$ simultaneously grow linearly, the I$'$ KK mode $\alpha$-vectors slide so that the type I (or I$'$) string theory oscillators provide the necessary light towers for the emergent string limits, and for these geodesics
\begin{align}
	\alpha_\parallel \geq 1/\sqrt{d-2}.
\end{align}

In these strong-coupling limits, several conjectures are satisfied. For instance, the above equations imply that the Sharpened Distance Conjecture is satisfied as one travels along geodesics into strong coupling. Also, the Weak Tower Alignment Conjecture, and possibly also the Light Tower Alignment Conjecture, hold. Additionally, in these limits, the $\alpha$-vectors of \eqref{e.16alphas} contain the ball of radius $1/\sqrt{d-2}$, and thus the Tower SWGC is satisfied.  Finally, the Heavy Tower Conjecture is satisfied in these limits because the $1/2$ BPS states become heavy as their $\alpha$-vectors are sufficiently anti-aligned with these rays and satisfy
\begin{align}
	|\alpha^\text{$1/2$ BPS}_\parallel|\geq 1/\sqrt{d-2}.
\end{align}

Meanwhile, for geodesics going to weak-coupling limits, some things are also quickly apparent. For instance, the $1/2$ BPS towers become light in these limits, as their $\alpha$-vectors in \eqref{e.16alphas} have projections $\alpha_\parallel\geq 1/\sqrt{d-2}$ along rays pointing in strong-coupling directions. This results in the Sharpened Distance Conjecture, as well as the Weak Tower Alignment Conjecture, and possibly also the Light Tower Alignment Conjecture.

Absent an understanding of heavy non-BPS states, it is difficult to fully test all of the conjectures from Section \ref{s.conjectures}. For instance, it is difficult to test to what extent the Tower Alignment Conjecture holds, because this requires non-BPS towers to align everywhere along the geodesics, and because geodesics here are straight lines, this conjecture needs that these non-BPS towers exist and align in both strong and weak coupling regimes. But here we have investigated these non-BPS towers only in regimes where they are light, thus testing only the Weak (and possibly Light) Tower Alignment Conjecture. These heavy towers are also necessary for testing the Tower SWGC in regimes where there is weak coupling. It is possible that such towers do exist, but it is not yet clear whether they do.

Though not considered in this example in this paper, it is interesting to consider geodesics where the axions vary as well. One might worry that the Tower Alignment Conjecture implies that there exist towers that violate emergent string limits by becoming lighter than emergent string towers.\footnote{I thank Alek Bedroya for drawing attention to this possibility.} To understand such a concern, consider two different geodesics that go to the same emergent string limit. Suppose that one of these geodesics, $\gamma_1$, is the geodesic that goes the direct route, and the fundamental string oscillators have $\alpha$-vectors parallel to $\gamma_1$. However, suppose that another distinct geodesic $\gamma_2$ goes indirectly to the same infinite distance limit by winding around the axionic directions. The Tower Alignment Conjecture says that there is a tower with a projection along $\gamma_2$ satisfying $\alpha_\parallel\geq 1/\sqrt{d-2}$. Since $\gamma_2$ is going to the same infinite distance limit, one might worry that this second tower becomes exponentially lighter than the direct fundamental string oscillator tower, since one might worry that this second tower travels an infinitely longer distance to get to the same infinite distance limit because of the winding.

It is likely that the resolution to this puzzle is that the different geodesics that go to the same emergent string limit differ by only a finite amount of length. This is because the moduli spaces involving these axionic directions are not flat, and a story similar to the IIB in 10d case likely applies here. As in the 10d IIB case, inequivalent infinite-distance geodesics likely only pass through the axionic cycles a few times before becoming parallel to direct infinite-distance geodesics. Thus, the Tower Alignment Conjecture likely does not contradict the Emergent String Conjecture in this example. Demonstrating this explicitly is the subject of future work.

\section{Conclusions and future directions\label{s.summary}}
In this paper, I have examined in examples how both the Tower SWGC and the Sharpened Distance Conjecture are both true but do not imply each other. My investigation has focused in particular on cases where the moduli spaces are not flat and the $\alpha$-vectors have moduli-dependence. I have found that the Tower SWGC and Sharpened Distance Conjecture follow from conjectures about dense infinite-distance geodesics and $\alpha$-vectors that align with these geodesics. I have also found other related conjectures.

Motivated by maximal supergravity examples, I have found evidence for and discussed connections between the following five conjectures:
\begin{enumerate}
	\item Dense Direction Conjecture.
	\item Tower Alignment Conjecture.
	\item Tower Scalar Weak Gravity Conjecture.
	\item Stringy Scalar Weak Gravity Conjecture.
	\item Heavy Tower Conjecture.
\end{enumerate}
In particular, I have discussed how Tower SWGC and Sharpened Distance Conjecture follow from the Dense Direction and Tower Alignment Conjectures simultaneously holding. I have also discussed how the Heavy Tower Conjecture can be viewed as a natural consequence of the Tower SWGC. I have also discussed possible strengthened and weakened versions of conjectures, and in particular I have in 10d cases used the Stringy SWGC as a proxy for the Tower SWGC.

It is not yet clear to what extent the conjectures of this paper apply beyond maximal supergravity. In Section \ref{s.169d}, I performed limited tests of these conjectures in half-maximal supergravity. Fuller tests require an understanding of non-BPS states that are heavy or in strongly-coupled regimes. Additionally, my 9d half-maximal supergravity tests did not involve axions, as I studied only the radion-dilaton submanifold of moduli space. It would be very beneficial to study in-depth how the Emergent String Conjecture, Sharpened Distance Conjecture, and Tower SWGC are satisfied in 9d theories with half-maximal supergravity.

In 10d maximal supergravity, I have tested the Stringy SWGC as a proxy for the Tower SWGC, and I have acted as if string oscillator towers exist and scale with the square roots of tensions of strings, even in non-perturbative regimes. When the strings in the Stringy SWGC oscillate with towers that scale with the square roots of tensions of the strings, then the Tower SWGC follows from the Stringy SWGC. But it is not clear how strongly-coupled strings oscillate, as perturbative string theory does not describe non-perturbative oscillations and potentially a semiclassical analysis is needed.

It might be interesting to study further the Stringy SWGC, and perhaps the Stringy SWGC can and should be generalized to a Membrane SWGC. A Membrane SWGC is potentially a necessary ingredient for the SWGC to be preserved under dimensional reduction, as the SWGC under dimensional reduction must involve branes and is still not yet fully understood. Also, often membranes are BPS, so they are under control and their $\alpha$-vectors can be studied everywhere in the moduli space. It is possible that there are many more things to say about $\alpha$-vectors for BPS membranes. Perhaps $\alpha$-vectors for BPS membranes could be used to locally distinguish geodesics that go to infinite distance limits from those that do not.

The Tower SWGC and the Tower Alignment Conjecture might be too strong, because they often require heavy (and possibly unstable) towers. If the Tower Alignment Conjecture is false, possibly only the Light or Weak Tower Alignment Conjectures hold. But, unlike the Tower Alignment Conjecture, the Light and Weak Tower Alignment Conjectures do not combine with the Dense Direction Conjecture to imply the Tower SWGC. It would be beneficial to obtain more evidence for, or against, the Tower SWGC and Tower Alignment Conjecture.

I have discussed how the Dense Direction Conjecture and Tower Alignment Conjecture imply the Sharpened Distance Conjecture and Tower SWGC, but the Emergent String Conjecture is not implied by my conjectures. It would be very interesting if somehow these conjectures could be strengthened so that the Emergent String Conjecture is implied. For instance, one might insist that, in the Tower Alignment Conjecture, towers that are not string oscillators have a KK-like tower spacing with masses $m_n\sim n$. This would result in some properties, but not necessarily all, of decompactification limits. A similar statement might be able to be made about emergent string limits. It would be interesting if these sorts of statements could be strengthened to imply the Emergent String Conjecture. This could be done by exploring in depth how emergent string limits arise in the simple examples considered in this paper.

It would be interesting to investigate the proposals with moduli spaces with boundaries. If the boundaries of these moduli spaces have reflexive boundary conditions, then the Dense Direction Conjecture could still apply in those examples.

It is possible that the conjectures of this paper apply to moduli spaces of quantum field theories. The Distance Conjecture has been studied in QFT contexts \cite{Perlmutter:2020buo}, but perhaps other Swampland conjectures such as the Tower SWGC, the Dense Direction Conjecture, and the Tower Alignment Conjecture also apply to such theories.

In this paper, the examples studied have been without potentials. Gradient flows of potentials have been connected with geodesics in \cite{Calderon-Infante:2022nxb}, so it is possible that $\alpha$-vectors and potentials are related. Perhaps some $\alpha$-vectors will be found to align with valleys of potentials.

The conjectures of this paper might nontrivially combine\footnote{Already, the Light Tower Alignment Conjecture makes specific references to the moduli-dependent species scale.} with the moduli-dependent species scale and the Desert Regions of \cite{Long:2021jlv, vandeHeisteeg:2022btw, vandeHeisteeg:2023ubh}.  For instance, in \cite{Rudelius:2023mjy} Rudelius proposed placing upper bounds on the lightest towers at each point in moduli space, and in examples he used the Desert Region to find these bounds. The Tower SWGC, together with properties of geodesics, might be able to shed light on these proposals. There is also the upcoming work of \cite{madridSpeciesHullTBA} that connects the species scale with $\alpha$-vectors and the Tower SWGC.

The convex hulls of $\alpha$-vectors often involve rotations of polytopes. These polytopes often have interesting rules governing the sizes and shapes of their facets, and a partial list of rules, along with a partial classification of the resulting polytopes, will appear in \cite{Etheredge:Taxonomy}.  But much about these shapes remains mysterious. Are these convex hulls related to deep mathematical questions? What rules fully classify these shapes?

Much remains to be understood about $\alpha$ vectors. They have played a prominent role in this paper, and they connect many Swampland Conjectures. For instance, the preservation of the Tower SWGC under dimensional reduction needs membranes and likely\footnote{The conditions necessary for the preservation of the Tower SWGC under dimensional reduction are still not yet fully understood.} the Weak Gravity Conjecture \cite{Etheredge:Bitowers}, thus potentially unifying several concepts. It is likely that $\alpha$-vectors will be used to construct grand and powerful conjectures that unify and sharpen many Swampland Conjectures.

\section*{Acknowledgements}
I am especially grateful to my PhD advisor Ben Heidenreich for his mentorship. For feedback on the manuscript, I am grateful to Stephanie Baines, Gina Etheredge, Ben Heidenreich, Yue Qiu, Matthew Reece, Tom Rudelius, Ignacio Ruiz, and Marin Sklan. I have benefited from discussions with Philip Argyres, Stephanie Baines, Jacob McNamara, Yue Qiu, Muthusamy Rajaguru, Sebastian Rauch, Matthew Reece, Elias Riedel G\aa rding, Tom Rudelius, Ignacio Ruiz, Cumrun Vafa, Irene Valenzuela, and David Wu. Part of this work was done while at TASI, and part of this work was done during the 2023 Simons Summer Workshop at the Simons Center for Geometry and Physics, Stony Brook University. I received support from the NSF grant PHY-2112800.

\appendix

\bibliographystyle{JHEP}
\bibliography{ref}
\end{document}